\documentclass[10pt,journal,compsoc]{IEEEtran}

\ifCLASSINFOpdf
\else
\fi

\usepackage{algorithmic}
\usepackage{array}
\usepackage{textcomp}
\usepackage{stfloats}
\usepackage{verbatim}
\usepackage{graphicx}
\usepackage{booktabs}
\usepackage{fleqn}
\usepackage{amssymb}
\usepackage{amsmath,amsfonts}
\usepackage{amsthm}
\usepackage{graphicx}
\usepackage{cite}
\usepackage{array}
\usepackage{color}
\usepackage{bm}
\usepackage{subfigure}
\usepackage{makecell}
\usepackage[linesnumbered,algoruled,boxed,lined,noend]{algorithm2e}
\usepackage{tabularx}
\usepackage{enumerate}
\usepackage{enumitem}
\usepackage[normalem]{ulem}
\usepackage{url}
\usepackage{microtype}
\usepackage{balance}
\usepackage{titlesec} 
\usepackage{mathrsfs}
\usepackage{tabularx}
\usepackage{multirow}
\usepackage{ulem}
\usepackage{bm}
\usepackage{pifont}

\titlespacing{\section}{0pt}{1.3ex}{1.3ex}
\titlespacing{\subsection}{0pt}{1.3ex}{1.3ex}
\titlespacing{\subsubsection}{0pt}{1.3ex}{1.3ex}

\allowdisplaybreaks[4]
\begin{document}

\title{Low-altitude Multi-UAV-assisted Data Collection and Semantic Forwarding for Post-Disaster Relief}

\author{Xiaoya Zheng,
        Geng Sun,~\IEEEmembership{Senior Member,~IEEE,}
        Jiahui Li,~\IEEEmembership{Member,~IEEE,}
        Jiacheng Wang, \\
        Weijie Yuan,~\IEEEmembership{Member,~IEEE,}
        Qingqing Wu,~\IEEEmembership{Senior Member,~IEEE,}
        Dusit Niyato,~\IEEEmembership{Fellow,~IEEE,} \\
        and Abbas Jamalipour,~\IEEEmembership{Fellow,~IEEE}
    
        \thanks{
        \par Xiaoya Zheng and Jiahui Li are with the College of Computer Science and Technology, Jilin University, Changchun 130012, China (E-mails: \protect\url{xiaoya257248@foxmail.com}; \protect\url{lijiahui0803@foxmail.com}).
        \par Geng Sun is with the College of Computer Science and Technology, Jilin University, Changchun 130012, China, and also with the Key Laboratory of Symbolic Computation and Knowledge Engineering of Ministry of Education, Jilin University, Changchun 130012, China. He is also with the College of Computing and Data Science, Nanyang Technological University, Singapore 639798 (E-mail: \protect\url{sungeng@jlu.edu.cn}).
        \par Jiacheng Wang and Dusit Niyato are with the College of Computing and Data Science, Nanyang Technological University, Singapore 639798 (E-mail: \protect\url{jiacheng.wang@ntu.edu.sg}; \protect\url{dniyato@ntu.edu.sg}).
        \par Weijie Yuan is with the Department of Electrical and Electronic Engineering, Southern University of Science and Technology, Shenzhen, Guangdong 518055, China (e-mail: \protect\url{yuanwj@sustech.edu.cn}).
        \par Qingqing Wu is with the Department of Electronic Engineering, Shanghai Jiao Tong University, Shanghai 200240, China (E-mail: \protect\url{qingqingwu@sjtu.edu.cn}).
        \par Abbas Jamalipour is with the School of Electrical and Computer Engineering, University of Sydney, Australia, and with the Graduate School of Information Sciences, Tohoku University, Japan (E-mail: \protect\url{a.jamalipour@ieee.org}).
        \par (Corresponding authors: Geng Sun and Jiahui Li). 
        }
}

\IEEEtitleabstractindextext{

\begin{abstract}
The low-altitude economy (LAE) is an emerging economic paradigm which fosters integrated development across multiple fields. As a pivotal component of the LAE, low-altitude uncrewed aerial vehicles (UAVs) can restore communication by serving as aerial relays between the post-disaster areas and remote base stations (BSs). However, conventional approaches face challenges from vulnerable long-distance links between the UAVs and remote BSs, and data bottlenecks arising from massive data volumes and limited onboard UAV resources. In this work, we investigate a low-altitude multi-UAV-assisted data collection and semantic forwarding network, in which multiple UAVs collect data from ground users, form clusters, perform intra-cluster data aggregation with semantic extraction, and then cooperate as virtual antenna array (VAAs) to transmit the extracted semantic information to a remote BS via collaborative beamforming (CB). We formulate a data collection and semantic forwarding multi-objective optimization problem (DCSFMOP) that jointly maximizes both the user and semantic transmission rates while minimizing UAV energy consumption. The formulated DCSFMOP is a mixed-integer nonlinear programming (MINLP) problem that is inherently NP-hard and characterized by dynamically varying decision variable dimensionality. To address these challenges, we propose a large language model-enabled alternating optimization approach (LLM-AOA), which effectively handles the complex search space and variable dimensionality by optimizing different subsets of decision variables through tailored optimization strategies. Simulation results demonstrate that LLM-AOA outperforms AOA by approximately 26.8\% and 22.9\% in transmission rate and semantic rate, respectively. Moreover, we reveal that the UAVs tend to be deployed near the center of the monitored area while avoiding excessive spatial concentration, thereby striking a balance between maintaining high-quality service for ground users and improving the efficiency of cluster-based collaborative transmission to the remote BS.
\end{abstract}

}
\maketitle

\IEEEdisplaynontitleabstractindextext
\IEEEpeerreviewmaketitle

%
%
\section{Introduction}
\label{sec: Introduction}

\noindent The low altitude economy (LAE) is an emerging economic paradigm~\cite{jiang20236gLAE}, which is primarily characterized by the deployment of aircraft as operational entities and is driven by diverse low-altitude flight activities. LAE fosters the development of multiple fields by enabling diverse services, including logistics, urban management, post-disaster emergency rescue, and tourism~\cite{Qin2025LAE},~\cite{ahmed2025toward}. Among these applications, post-disaster emergency rescue is particularly critical, which encompasses disaster reconnaissance and assessment, provision of temporary communication services, delivery of emergency supplies, and infrastructure repair.

\par As a core component of the emerging LAE, low-altitude uncrewed aerial vehicles (UAVs) play a vital role in post-disaster communications~\cite{wang2022task},~\cite{wang2023secure}. Equipped with portable base stations (BSs) and sensing payloads, UAVs can be rapidly deployed to collect data from isolated ground users when terrestrial infrastructure is damaged~\cite{wan2024deep}. In such scenarios, UAVs must be properly positioned to establish reliable and efficient air-to-ground links under stringent onboard energy constraints, which poses significant challenges for UAV deployment and user association.

\par After completing the data collection, UAVs must forward the aggregated data to a remote BS for further processing. However, data forwarding in UAV-assisted post-disaster networks faces several critical challenges. \underline{\emph{First}}, when terrestrial infrastructure is incapacitated, UAVs are required to transmit data directly to the remote BS. However, severe path loss, blockage, and limited transmit power often result in insufficient signal strength and unstable communication links. \underline{\emph{Second}}, post-disaster applications generate large volumes of task-critical textual information, such as situation reports, survivor updates, and mission coordination messages, while the achievable link capacity is severely constrained. The mismatch renders efficient data delivery extremely challenging. \underline{\emph{Finally}}, the communication link between the UAVs and BS can be easily congested when sudden traffic surges, which necessitates novel transmission mechanisms that are capable of flexibly adapting the communication rate in response to dynamic network conditions and data priorities. Therefore, it is imperative to apply effective techniques to address the aforementioned challenges.

\par Collaborative beamforming (CB) provides an effective approach in addressing the first challenge. Specifically, based on the principle of spatial coherence, CB allows multiple UAVs to coordinate their transmission parameters to achieve constructive signal superposition in the desired direction, thereby forming a high-gain directional beam and suppressing undesired radiation~\cite{Mozaffari2019LAA},~\cite{Jayaprakasam2017}. Accordingly, CB exhibits the following notable advantages. \underline{\emph{First}}, signal superposition substantially enhances the received power, which alleviates the adverse effects of long propagation distances and obstructions in post-disaster environments. \underline{\emph{Second}}, CB improves transmission efficiency by concentrating energy toward the target BS, which in turn allows UAVs to allocate more resources to other rescue tasks. \underline{\emph{Finally}}, CB provides robustness against UAV failures, as the remaining UAVs can adapt their parameters to maintain the communication link. In conclusion, these advantages ensure robust and resilient post-disaster communications.

\par Semantic communication offers a feasible solution to the latter two challenges. Specifically, it extracts task-relevant meaning from raw data, removes redundant information, and compresses data while preserving essential semantics~\cite{xie2021deep},~\cite{chaccour2024less}. Integrating semantic communication with UAV-assisted post-disaster networks yields the following advantages. \underline{\emph{First}}, transmitting only semantic information alleviates the mismatch between limited link capacity and large data volumes, thereby enabling efficient delivery under bandwidth constraints. \underline{\emph{Second}}, the semantic rate can be adaptively adjusted by controlling the number of transmitted semantic symbols according to channel conditions, data urgency, and task requirements, which is well suited to dynamic post-disaster environments. \underline{\emph{Finally}}, the remote BS can directly perform high-level inference on refined semantic data with reduced processing overhead, thereby improving the timeliness and effectiveness of disaster response. These advantages highlight the potential of semantic communication in enhancing the efficiency and resilience of UAV-assisted post-disaster communication systems.

\par However, the realization of such a low-altitude UAV-assisted post-disaster network which integrates CB and semantic communication is inherently complex. \underline{\emph{First}}, the design of such a sophisticated architecture involves substantial engineering complexity. \underline{\emph{Second}}, even though the architecture is established, how to clarify the coupling relationships among the multiple decision variables of UAVs, including data acquisition, UAV clustering, semantic compression as well as collaborative transmission, and then formulate the corresponding optimization problem are challenging. \underline{\emph{Finally}}, developing efficient algorithms to jointly optimize these interdependent decision variables for enhanced system performance is nontrivial. These challenges motivate this work, and the main contributions are summarized as follows.

\begin{itemize}

\item \textbf{\emph{Post-disaster Low-altitude Multi-UAV-assisted Data Collection and Semantic Forwarding Communication Network:}} We consider a network where low-altitude UAVs initially collect data from their respective ground users and then autonomous clustering. Within each cluster, data are aggregated, semantically compressed, and transmitted to a remote BS via CB. The architecture is well-suited for post-disaster relief, since data aggregation and compression substantially reduce the transmitted data volume while CB enhances the transmission efficiency.

\item \textbf{\emph{Data Collection and Semantic Forwarding Muti-objective Optimization Problem:}} We formulate an optimization problem which aims to maximize the total transmission rate of all served users, maximize the total semantic transmission rate of all clusters, and minimize the flight energy consumption of all UAVs by joint optimizing UAV clustering, locations, excitation current weights, and the number of semantic symbols used per word in each cluster. The formulated problem is a mixed-integer nonlinear programming (MINLP) problem, inherently NP-hard, with dynamically varying decision dimensions.

\item \textbf{\emph{Large Language Model-enabled Alternating Optimization Approach:}} We propose a large language model-enabled alternating optimization approach (LLM-AOA) to solve the formulated problem by synergistically integrating greedy strategy and evolutionary computation. Specifically, the decision variables are decomposed into three components, where the UAV clustering is optimized by the greedy clustering algorithm (GCA), UAV locations and excitation weights are optimized by LLM-guided non-dominated sorting genetic algorithm II (NSGA-II), and the number of semantic symbols per word in each cluster is determined by greedy symbol optimization (GSO). By applying tailored optimization methods for each component based on its characteristics, LLM-AOA enables more effective exploration and exploitation of the solution space.
    
\item \textbf{\emph{Performance Validation of the Optimization Approach:}} Simulation results illustrate the effectiveness and superiority of the proposed LLM-AOA in improving the performance of the low-altitude multi-UAV-assisted data collection and semantic forwarding communication network. We find that UAVs tend to be located in the vicinity of the monitor area center, while avoiding overly dense distribution. The spatial distribution reflects a trade-off between ensuring high-quality service for ground users and improving the efficiency of cluster-based collaborative transmission to the remote BS.
    
\end{itemize}

\par The remainder of this work is organized as follows. Section \ref{sec: Related Works} reviews the related works. Section \ref{sec: System Model} introduces the system model. Section \ref{sec: Problem Formualtion and Analysis} formulates the optimization problem and presents the corresponding analysis. Section \ref{sec: Algorithm} proposes the approach. Section \ref{sec: simulation} provides the simulation results. Finally, Section \ref{sec: Conclusion} concludes the paper.

%
%
\section{Related Work}
\label{sec: Related Works}

\noindent In this section, we review the related studies on low-altitude UAV-assisted post-disaster communications, optimization metrics, and optimization approaches in low-altitude UAV-assisted wireless communications.

\subsection{Low-altitude UAV-assisted Post-disaster Communications}
\label{subsubsec: UAV-assisted Post-disaster Communications}

\noindent {\emph{Low-altitude Multi-UAV-enabled Data Collection.} Extensive research has investigated the use of multi-UAV-enabled data collection in post-disaster scenarios. For example, Wan \emph{et al.}~\cite{Wan2024multipleUAVdatacollection} analyzed route planning and service time allocation for multiple UAVs in the context of time-varying post-disaster data volume from each ground node. Lin \emph{et al.}~\cite{Lin2022multipleUAVdatacollection} proposed a multi-subcarrier non-orthogonal multiple access (NOMA)-based multi-UAV emergency communication network to collect messages from ground users. Subsequently, they introduced a new metric called prioritized delay and formulated an optimization problem to strike a balance between minimizing the maximum prioritized delay of all users and suppressing multi-user interference. Fu \emph{et al.}~\cite{Fu2025multipleUAV} constructed a heterogeneous UAV-based Internet of Things (IoT) system architecture in which the collector UAVs collect data from ground sensors and the follower UAVs are responsible for aggregating all the collected data and uploading it to a satellite. However, the aforementioned works focus solely on the transmission from ground users to UAV, while neglecting the link for transmitting the collected data to the remote, which may lead to suboptimal performance of the overall communication system.

\par \emph{Low-altitude Multi-UAV-enabled data forwarding.} Several studies have focused on data forwarding in multiple-UAV-enabled post-disaster communication networks. For example, Dai \emph{et al.}~\cite{Dai2022multipleUAVdatacollection} optimized the downlink channel allocation between the UAVs and ground users in a multi-UAV-assisted emergency communication system. Subsequently, based on the obtained channel allocation scheme, they proposed an optimal data delivery strategy by using a Stackelberg game to enhance the system transmission efficiency. Moreover, Zheng \emph{et al.}~\cite{Zheng2024TWC} investigated a UAV-based relay system in which the UAVs form virtual antenna arrays (VAAs) to transmit the collected data to multiple different BSs. Accordingly, they formulated a multi-objective optimization problem (MOP) to improve the overall system performance. However, the aforementioned studies primarily focus on optimizing transmissions for data delivery to remote BSs or command centers, while treating data collection as a predefined process. The decoupled designs neglect the intrinsic interdependence between data collection and forwarding in post-disaster scenarios. 

\subsection{Optimization Metrics in Low-altitude UAV-assisted Wireless Communications}
\label{subsubsec: Optimization Metrics in UAV-assisted Wireless Communications}

\noindent \emph{Optimization of Transmission Rate.} Several studies have investigated transmission rate maximization problems in UAV communications and networking. For example, Ajam \emph{et al.}~\cite{Ajam2020rate} investigated a UAV-assisted relay system, where the link between the UAV and ground users serves as a radio frequency (RF) access link, and the link between the UAV and BS functions as a free-space optical (FSO) backhaul link. Accordingly, they adopted the ergodic sum rate as the optimization metric to evaluate end-to-end system performance. Zhao \emph{et al.}~\cite{ZhaoNAN2019tcom} studied a NOMA-based communication system in which the UAV and BS collaboratively serve ground users and aimed to maximize the sum rate of users through optimal user scheduling and UAV trajectory design. Liu \emph{et al.}~\cite{Liu2019twc} proposed a multi-antenna UAV-BS coexisting system where the BSs are categorized into \emph{available GBSs} that do not serve terrestrial users and \emph{occupied GBSs} that provide services to their respective terrestrial users. They first characterized the degrees of freedom (DoF) and then maximized the corresponding sum-rate at finite SNR. Moreover, Yang \emph{et al.}~\cite{Yang2023TVT} analyzed the rate performance in cellular-enabled UAV swarm networks, formulated two typical optimization problems, \emph{i.e.}, sum rate maximization and max-min rate optimization, and tackled them by using Karush–Kuhn–Tucker and iterative algorithms, respectively.

\par \emph{Optimization of Semantic Rate.} Several studies have investigated semantic rate optimization in UAV-assisted semantic communication systems. For example, Liu \emph{et al.}~\cite{Liu2025TMC} integrated semantic communication with UAV-enabled mobile edge computing (MEC) systems under jamming attacks. Then, they optimized the semantic rate between the users and UAVs, along with other performance metrics, by jointly designing UAV trajectories, user associations, and channel selections. Xie \emph{et al.}~\cite{xie2025noma} proposed a NOMA-based multi-UAV semantic communication network in which each UAV transmits semantic data to multiple users and aimed to maximize the overall semantic data transmission rate. Moreover, Hu \emph{et al.}~\cite{Hu2025TCOM} incorporated the semantic rate as part of the quality-of-experience (QoE) metric in a UAV-enabled semantic communication system and applied deep reinforcement learning (DRL) to optimize it through UAV trajectory planning, spectrum bandwidth allocation, and transmit power design.

\par However, the aforementioned studies primarily focus on optimizing rate-related performance metrics while neglecting the essential factor of energy consumption. The oversight poses a significant limitation for practical UAV-assisted communication systems, where energy resources are inherently constrained.

\subsection{Optimization Approaches in Low-altitude UAV-assisted Wireless Communications}
\label{subsubsec: Optimization Methods in UAV-assisted Wireless Communications}

\noindent \emph{Convex Optimization Methods.} Several prior studies have employed convex optimization techniques in UAV communications and networking. For example, Wang \emph{et al.}~\cite{wang2024latency} proposed a UAV-assisted MEC network architecture and decomposed the formulated optimization problem into three subproblems, which were solved by using a block coordinate descent (BCD)-based algorithm. Sun \emph{et al.}~\cite{sun2023max} adopted successive convex approximation (SCA) techniques for trajectory design and transmission scheduling in solar-powered UAV systems. Zhang \emph{et al.}~\cite{Zhang2024convex} investigated the joint optimization in dynamic user multi-UAV MEC systems and applied both SCA as well as BCD methods to address resource allocation and trajectory design. Feng \emph{et al.}~\cite{feng2024optimal} studied a UAV-enabled IoT communication network employing rate-splitting multiple access (RSMA), where the data of each IoT is divided into common and private parts. They formulated a joint optimization problem to maximize system throughput, decomposed it into multiple subproblems via BCD, and proposed an alternating optimization (AO) scheme to obtain the final solution. Moreover, Zhou \emph{et al.}~\cite{Zhou2025convex} investigated a quality-of-service (QoS)-oriented NOMA-assisted UAV-MEC system and introduced a novel metric, \emph{i.e.}, the system overhead ratio (SOR), to quantify the communication QoS. To minimize the SOR, a hybrid Lyapunov–convex optimization-based scheme was proposed. However, the inherently non-convex nature of these optimization problems often necessitates complex reformulations to achieve convexity. Moreover, the dynamic dimensionality of decision variables during optimization hinders the direct application of conventional convex optimization methods.

\par \emph{DRL-based Optimization Methods.} Numerous studies have explored the application of deep reinforcement learning (DRL) techniques to address optimization problems in UAV communications and networking. For example, Tang \emph{et al.}~\cite{tang2024deep} investigated a UAV-assisted wireless-powered IoT network and employed the data-driven deep deterministic policy gradient (DDPG) algorithm to optimize three-dimensional (3D) UAV trajectories. Chen \emph{et al.}~\cite{Chen2025DRL} studied a UAV-enabled MEC system and proposed a scheme named JTORA, which integrates DRL and Lyapunov optimization to jointly optimize UAV trajectory and communication resources. Peng \emph{et al.}~\cite{Peng2024DRL} investigated the collaboration between the UAVs and intelligent reflecting surfaces (IRSs) to minimize the data harvesting time by jointly designing UAV trajectories and IRS parameters, where a memory-based softmax deep double deterministic policy gradient (DDPG) algorithm is used for trajectory optimization, and theoretical analysis is employed for IRS control. Moreover, multi-agent DRL (MADRL) methods have also been widely adopted for UAV-related optimization. For example, Qin \emph{et al.}~\cite{qin2023deep} deployed multiple UAVs for aerial integrated sensing and communications (ISAC) and adopted multi-agent soft actor-critic (MASAC) to address the decision-making problem. Moreover, Betalo \emph{et al.}~\cite{Betalo2025DRL} studied a laser-charged UAV-assisted rechargeable wireless sensor network, formulated a joint optimization problem to minimize mission completion time and node death time, and proposed a MADRL-based real-time charging scheduling scheme. However, both traditional single-agent and multi-agent DRL algorithms typically consider fixed-dimensional state and action spaces, which makes them unsuitable for scenarios where structural dependencies among decision variables lead to variable-dimensional optimization problems. 

\subsection{Main Difference from Other Studies}
\label{subsec: Main Difference from Other Works}

\noindent Different from the aforementioned related works, we investigate a post-disaster low-altitude multi-UAV-assisted data collection and semantic forwarding communication network in which the UAVs collect data from ground users, extract semantic features, and finally relay the key data to a remote BS. Accordingly, we formulate an optimization problem to enhance the communication performance while considering the energy consumption of UAVs. Moreover, to efficiently solve the formulated optimization problem, we develop an AO-based approach that iteratively optimizes different components of the decision variables.

%
%
\section{System Model}
\label{sec: System Model}

\noindent In this section, we first present the architecture of the low-altitude multi-UAV-assisted data collection and semantic forwarding communication network. Next, we describe the semantic communication model, data transmission model, and energy consumption model of UAVs. 

\begin{figure}
\centerline{\includegraphics[width=3.5in]{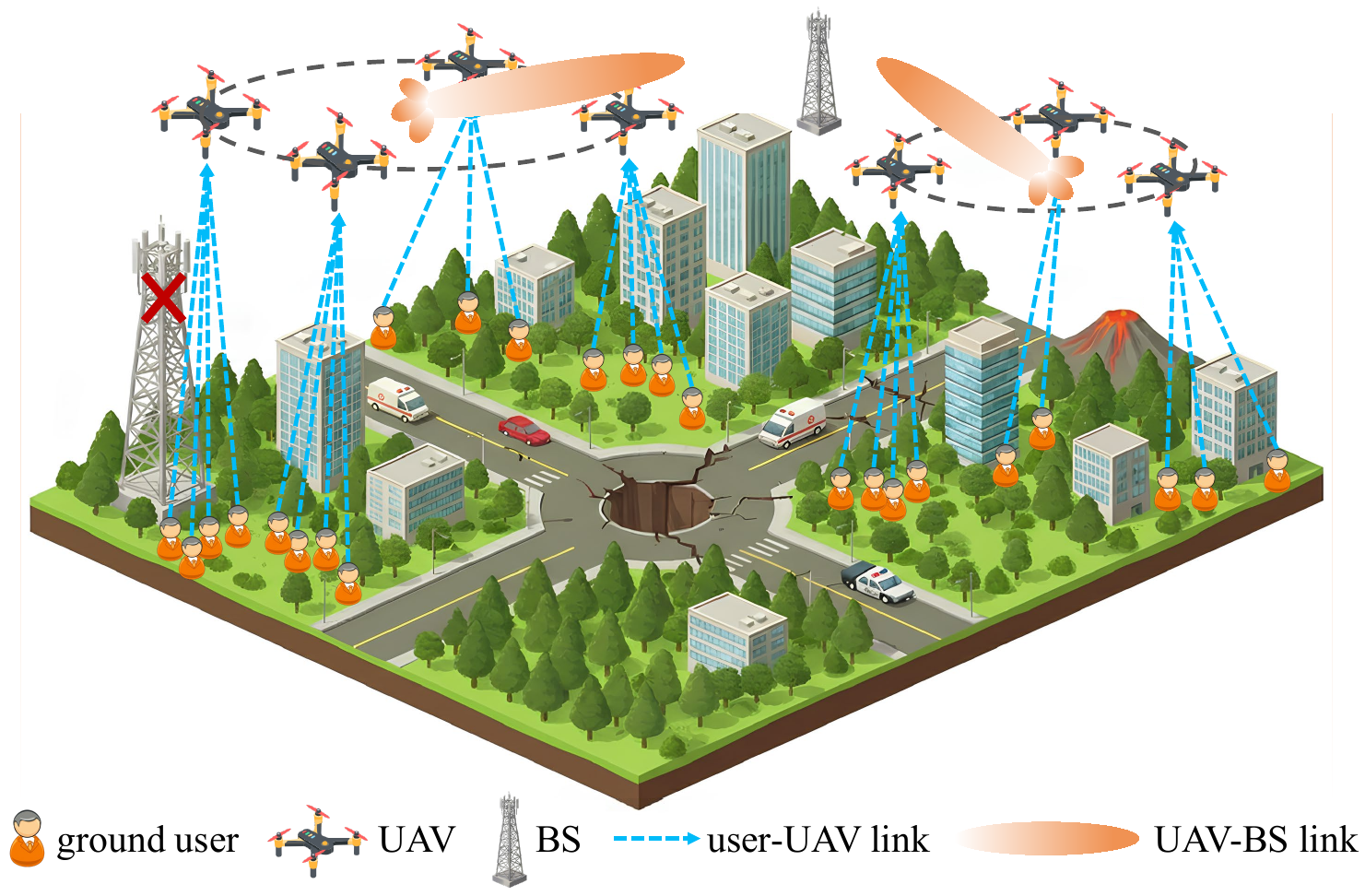}}
\caption{The low-altitude multi-UAV-assisted data collection and semantic forwarding communication network composed of multiple ground users, UAVs, and a remote BS.}
\label{fig. sketch map}
\end{figure}

\subsection{Network Architecture}
\label{subsec: Network Architecture}

\noindent Fig. \ref{fig. sketch map} illustrated the architecture of the low-altitude multi-UAV-assisted data collection and semantic forwarding communication network, which primarily comprises the following components:

\begin{itemize}

\item \emph{Ground Users.} Multiple ground users in the post-disaster area, denoted by $ \mathcal{U} = \{ u \mid 1 \leq u \leq N_{\mathrm{U}} \} $, need to transmit data to a remote BS. The data set of the $u$-th user is represented as $ \mathcal{S}_u $ $= \{ \mathbf{s}_{u, 1}, $ $ \dots, $ $\mathbf{s}_{u, d},$ $\dots, $ $\mathbf{s}_{u, D_u} \} $, where $ D_u $ denotes the total number of sentences of the $u$-th user. Each sentence $ \mathbf{s}_{u, d}$ $= [ w_{u, d, 1},$ $\dots, $ $w_{u, d, l},$ $\dots,$ $w_{u, d, L_{u, d}} ] $ consists of $ L_{u, d} $ words, and $ w_{u, d, l} $ denotes the $ l $-th word in the $ d $-th sentence of the $u$-th user.

\item \emph{UAVs.} Multiple UAVs equipped with the semantic communication model~\cite{xie2021deep} operate over the post-disaster area, denoted by $\mathcal{V} = \{v, 1 \leq v \leq N_{\mathrm{V}}\}$, where $N_{\mathrm{V}}$ represents the total number of UAVs. The $v$-th UAV provides wireless communication services for a portion of ground users, denoted by $\mathcal{U}_{v}$. The UAVs self-organize into $N_{\mathrm{cluster}}$ clusters, denoted by $\mathscr{C} = \{\mathcal{C}_{1}, \ldots, \mathcal{C}_{c}, \ldots, \mathcal{C}_{N_{\mathrm{cluster}}}\}$. Within the $c$-th cluster, the UAVs are represented by $\mathcal{V}_{c} = \{v_{c}, 1 \leq v_{c} \leq N_{\mathrm{V}_{c}}\}$, where $N_{\mathrm{V}_{c}}$ denotes the number of UAVs in the $c$-th cluster. 

\item \emph{BS.} A remote BS is responsible for receiving data from the post-disaster area.

\end{itemize}

\par Accordingly, the operation mechanism of the network for post-disaster communications is presented as follow:

\begin{enumerate}[label=\alph*)]

\item The $u$-th user transmits its data $\mathcal{S}_{u}$ to the nearest UAV within its communication range. During this stage, a UAV is capable of simultaneously receiving data from multiple users through multiplexing techniques.

\item The $v_c$-th UAV aggregates the collected data, which is represented as $\mathcal{S}_{v_c} = \bigcup_{u \in \mathcal{U}_{v}} \mathcal{S}_{u}$, and then engages in intra-cluster data sharing. During this period, the set of sentences collected by the $c$-th UAV cluster is represented as $\mathcal{S}_{c} = \bigcup_{v_c \in \mathcal{V}_{c}} \mathcal{S}_{v_c} = \bigcup_{v_c \in \mathcal{V}_{c}} \bigcup_{u \in \mathcal{U}_{v_c}} \mathcal{S}_{u}$.

\item All UAVs in the $c$-th cluster extract semantic feature by using the equipped semantic communication model and then transmit the extracted data to the remote BS through CB.

\end{enumerate}

\par We consider a 3D Cartesian coordinate system where the locations of the $u$-th ground user, the $v$-th UAV in the $c$-th cluster, and the remote BS are denoted by $\textbf{q}_{u} = [x_{u}, y_{u}, z_{u}]$, $\textbf{q}_{v_{c}} = [x_{v_{c}}, y_{v_{c}}, z_{v_{c}}]$, and $\textbf{q}_{\mathrm{bs}} = [x_{\mathrm{bs}}, y_{\mathrm{bs}}, z_{\mathrm{bs}}]$, respectively.

\subsection{Semantic Communication Model}
\label{subsec: DeepSC Model}

\noindent We consider that each UAV is equipped with the well-known DeepSC model~\cite{Xie2021DeepSC}. During the DeepSC-based communication process, the transmitter first encodes the collected data into semantic symbols, which are then transmitted through the physical channel and subsequently decoded by the receiver to reconstruct the original information. The three components of DeepSC are described as follows.

\par \emph{1) Transmitter.} Given the data set $\mathcal{S}_{c}$ collected by the $c$-th UAV cluster, the DeepSC transmitter performs semantic encoding. The resulting symbol stream is as follows~\cite{Xie2021DeepSC}:
\begin{equation}
\label{eq. encoded symbol stream}
\mathbf{x}_{c} = \mathbf{C}_{\bm{\beta}}(\mathbf{S}_{\bm{\alpha}}(\mathcal{S}_{c})),
\end{equation}
\noindent where $\mathbf{S}_{\bm{\alpha}}$ and $\mathbf{C}_{\bm{\beta}}$ are the semantic encoder and the channel encoder, respectively.

\par \emph{2) Physical Channel.} The encoded symbols $\mathbf{x}_{c}$ are transmitted to the receiver via a fading channel, which is modeled as follows~\cite{Xie2024semantic}:
\begin{equation}
\label{eq. physical channel}
\mathbf{y}_{c} = h\mathbf{x}_{c} + \mathbf{n},
\end{equation}
\noindent where $h$ denotes the channel coefficient and $\mathbf{n}$ is the additive white Gaussian noise (AWGN).

\par \emph{3) Receiver.} On the receiver side, the received signal $\mathbf{y}_{c}$ is successively processed to recover the transmitted data, which is as follows~\cite{Yan2022semantictext,Xie2024semantic}:
\begin{equation}
\label{eq. decoded symbol stream}
\hat{\mathcal{S}}_{c} = \mathbf{S}^{-1}_{\bm{\chi}}(\mathbf{C}^{-1}_{\bm{\psi}}(\mathbf{y}_{c})),
\end{equation}
\noindent where $\mathbf{C}^{-1}_{\bm{\psi}}$ and $\mathbf{S}^{-1}_{\bm{\chi}}$ denote the channel decoder and semantic decoder parameterized by $\bm{\psi}$ and $\bm{\chi}$, respectively.

\subsection{Data Transmission Model}
\label{subsec: Data Transmission Model}

\noindent In this subsection, we first present the VAA and channel models, followed by the characterization of end-to-end data transmission from ground users to UAVs and from UAV clusters to the remote BS.

\subsubsection{Virtual Antenna Array Model}
\label{subsubsec: VAA Model}

\noindent We use the array factor (AF) to characterize the beam pattern of each VAA. Specifically, the AF of the VAA formed by UAVs in the $c$-th cluster is as follows~\cite{Li2021}:
\begin{equation}
\small
\label{eq. AF}
F_c(\theta,\phi)=\sum\nolimits_{v=1}^{N_{\mathrm{V}_{c}}}w_{v_{c}}\text{e}^{j[p(x_{v_{c}}\sin\theta\cos\phi + y_{v_{c}}\sin\theta\sin\phi + z_{v_{c}}\cos\theta)]},
\end{equation}
\noindent where $w_{v_{c}}$ is the excitation current weight of the $v$-th UAV in the $c$-th cluster, $j$ represents the imaginary unit, and $p = 2\pi/\lambda$ denotes the phase constant with $\lambda$ being the carrier wavelength. Moreover, $\theta \in [0, \pi]$ and $\phi \in [-\pi, \pi]$ denote the elevation and azimuth angles, respectively.

\par Accordingly, the antenna gain $G_{c}$ of the VAA formed by UAVs in the $c$-th cluster is as follows~\cite{Mozaffari2019LAA}:
\begin{equation}
\label{eq. G}
G_{c} = \frac{4\pi|F_{c}(\theta_{c, \mathrm{bs}},\phi_{c, \mathrm{bs}})|^2w(\theta_{c, \mathrm{bs}},\phi_{c, \mathrm{bs}})^2}{\int_0^{2\pi}\int_0^{\pi}|F_{c}(\theta,\phi)|^2w(\theta,\phi)^2\sin\theta\,d\theta\,d\phi}\eta,
\end{equation}
\noindent where $(\theta_{c, \mathrm{bs}},\phi_{c, \mathrm{bs}})$ is the direction of the remote BS, $w(\theta,\phi)$ represents the far-field beam pattern magnitude of each UAV, and $\eta$ denotes the array efficiency.

\subsubsection{Channel Model}
\label{subsubsec: Channel Model}

\noindent The links between the ground users and UAVs, as well as those between the UAV clusters and remote BS, follow a probabilistic line-of-sight (LoS) model. Specifically, the LoS probability between the $u$-th ground user and the $v$-th UAV is expressed as~\cite{AlHourani2014}:
\begin{equation}
\label{eq. LoS probability}
P^{\mathrm{LoS}}_{u,v} = \frac{1}{1 + \psi \exp \!\left[-\beta \left(\frac{180}{\pi} \arcsin \!\left(\frac{H_{u,v}}{d_{u,v}}\right) - \psi \right)\right]},
\end{equation}
\noindent where $\psi$ and $\beta$ are environment-dependent constants. Moreover, $H_{u,v}$ and $d_{u,v}$ represent the vertical and total distances between the $u$-th ground user and the $v$-th UAV, respectively. Accordingly, the non-LoS (NLoS) probability is $P^{\mathrm{NLoS}}_{u,v} = 1 - P^{\mathrm{LoS}}_{u,v}$.

\par The path loss between the $u$-th ground user and the $v$-th UAV for LoS and NLoS links is as follows~\cite{AlHourani2014}:
\begin{equation}
L_{u,v}^{\text{LoS}} = 20 \log d_{u,v} + 20 \log f + 20 \log \left( 4 \pi/c \right) + \mu_{\text{LoS}},
\end{equation}
\begin{equation}
L_{u,v}^{\text{NLoS}} = \underbrace{20 \log d_{u,v} + 20 \log f + 20 \log \left( 4 \pi/c \right)}_{\text{free space path loss}} + \underbrace{\mu_{\text{NLoS}}}_{\substack{\text{excessive} \\ \text{path loss}}},
\end{equation}
\noindent where $\mu_{\mathrm{LoS}}$ and $\mu_{\mathrm{NLoS}}$ denote the excess path loss coefficients for LoS and NLoS links, respectively. Moreover, $f$ is frequency and $c$ represent light speed.

\par Accordingly, the average path loss between the $u$-th ground user and the $v$-th UAV is as follows~\cite{AlHourani2014}:
\begin{equation}
\label{eq. average channel power gain between user and UAV}
L_{u, v} = \mathit{P}^{\mathrm{LoS}}_{u, v}L_{u,v}^{\text{LoS}} + \mathit{P}^{\mathrm{NLoS}}_{u, v}L_{u,v}^{\text{NLoS}}.
\end{equation}

\par Note that the calculation of the average path loss $L_{c, \mathrm{bs}}$ between the $c$-th UAV cluster and remote BS follows the same procedure as described above.

\subsubsection{Transmission Model}
\label{subsubsec: Transmission Model}

\noindent In this part, we elaborate the transmission models for both ground user-to-UAV and UAV cluster-to-BS links.

\par \emph{1) Ground User-to-UAV Transmission Model.} Each ground user may fall within the coverage areas of multiple UAVs. For simplicity, we assume that each ground user establishes a communication link with the geographically closest UAV. Specifically, the $v_u^\star$-th UAV which is associated with the $u$-th ground user 
is given as follows:
\begin{equation}
v_u^\star = \arg\min_{v \in \mathcal{V}} d_{u,v}.
\end{equation}

\par Accordingly, the SINR between the $u$-th ground user and the $v_u^\star$-th UAV is as follows~\cite{ma2022robustrate}:
\begin{equation}
\text{SINR}_{u, v_u^\star} = \frac{P_u 10^{-\frac{L_{u, v_u^\star}}{10}}}{\sum_{u' \in \mathcal{U}_{v_u^\star} \setminus \{u\}} P_{u'} 10^{-\frac{L_{u', v_u^\star}}{10}} + BN_0},
\label{eq:SINR_uv}
\end{equation}
\noindent where $P_u$ is the transmit power of the $u$-th ground user, $B$ denotes the channel bandwidth, and $N_0$ is the noise power spectral density. Moreover, $\mathcal{U}_{v_u^\star}$ denotes the set of ground users served by the $v_u^\star$-th UAV.

\par Accordingly, the achievable transmission rate between the $u$-th user and the selected $v_u^\star$-th UAV is as follows~\cite{Zeng2017efficiency}:
\begin{equation}
R_{u, v_u^\star} = B \log_2 \left(1 + \text{SINR}_{u, v_u^\star} \right).
\label{eq:rate_uv}
\end{equation}

\par \emph{2) UAV Cluster-to-BS Transmission Model.} The BS is assumed to be equipped with multiple antennas, which enables it to distinguish signals originating from different UAV clusters. Consequently, the signal-to-noise ratio (SNR), rather than the SINR, is considered as the performance metric for the link between each cluster and BS. Specifically, the SNR between the $c$-th UAV cluster and BS is as follows~\cite{Zheng2024SNR}:
\begin{equation}
\text{SNR}_{c, \mathrm{bs}} = 
\begin{cases}
\frac{P_{c} G_{c} 10^{-\frac{L_{c, \textrm{bs}}}{10} }}{BN_0}, & \text{if } N_{\mathrm{V}_{c}} \neq 1, \\
\frac{P_{v} 10^{-\frac{L_{c, \textrm{bs}}}{10} }}{BN_0}, & \text{if } N_{\mathrm{V}_{c}} = 1,
\end{cases}
\label{eq. SNR_cs}
\end{equation}
\noindent where $P_{c}$ is the total transmit power of the $c$-th UAV cluster, given by $P_{c} = \sum \nolimits_{v=1}^{N_{\mathrm{V}_{c}}}(w_{v_{c}})^2 P_{v}$, with $P_{v}$ being the transmit power of a single UAV. Accordingly, the semantic transmission rate between the $c$-th UAV cluster and BS is as follows~\cite{Liu2024semanticrate}:
\begin{equation}
\text{SR}_{c, \mathrm{bs}} = \frac{BI}{k_{c, \mathrm{bs}}L}\xi_{c, \mathrm{bs}} \{k_{c, \mathrm{bs}}, \text{SNR}_{c, \mathrm{bs}}\},
\label{eq: rate_cs}
\end{equation}
\noindent where $I$ and $L$ represent the expected number of semantic information and words per sentence, respectively. Moreover, $k_{c, \mathrm{bs}}$ denotes the average number of semantic symbols used per word in the $c$-th cluster. In addition, $\xi_{c, \mathrm{bs}}$ is the resulting semantic similarity between the $c$-th UAV cluster and BS, which is determined as a function of both $k_{c, \mathrm{bs}}$ and $\text{SNR}_{c, \mathrm{bs}}$.

\subsection{Energy Consumption Model}
\label{subsec: Energy Consumption Model}

\noindent In this subsection, we present the flight energy consumption of UAVs. Specifically, for the $v$-th UAV in the $c$-th cluster, flying at a horizontal speed of $V_{v_{c}}^{\textrm{xy}}$, the propulsion power consumption is as follows~\cite{Zeng2019energyminimization},~\cite{Meng2022energycons}:
\begin{align}
P_{v_{c}}^{\textrm{xy}}(V_{v_{c}}^{\textrm{xy}}) = & \underbrace{P_{\mathrm{ind}} \left(\sqrt{1 + \frac{\|V_{v_{c}}^{\textrm{xy}}\|^{4}}{4V_{\mathrm{ind}}^{4}}}-\frac{\|V_{v_{c}}^{\textrm{xy}}\|^{2}}{2V_{\mathrm{ind}}^{2}} \right)^{1/2}}_{\text{induced}} + \nonumber  \\ 
& \underbrace{P_{\mathrm{0}}\left(1+\frac{3 \|V_{v_{c}}^{\textrm{xy}}\|^{2}}{V_{\mathrm{tip}}^{2}}\right)}_{\text{blade profile}} + \underbrace{\frac{1}{2} d_0 \rho sA \|V_{v_{c}}^{\textrm{xy}}\|^3}_{\text{parasite}}, 
\label{eq. horizontal power}
\end{align}
\noindent where $P_{\mathrm{ind}}$ and $P_{\mathrm{0}}$ are the induced power during hovering and the blade profile power, respectively. Moreover, $V_{\mathrm{tip}}$ and $V_{\mathrm{ind}}$ are the rotor tip speed and the mean induced velocity in hovering, respectively. In addition, $d_0$, $\rho$, $s$, and $A$ correspond to the fuselage drag ratio, air density, rotor solidity, and rotor disk area, respectively.

\par Moreover, for the $v$-th UAV in the $c$-th cluster flying with speed $V_{v_{c}}^{\textrm{z}}$ vertically, the propulsion power consumption is as follows~\cite{ZhangChuang2025TMC}:
\begin{equation}
P_{v_{c}}^{\textrm{z}}(V_{v_{c}}^{\textrm{z}}) = 
\begin{cases}
W V_{v_{c}}^{\textrm{z}}, &  V_{v_{c}}^{\textrm{z}} > 0,\\
0, & V_{v_{c}}^{\textrm{z}} \leq 0,
\end{cases}
\label{eq. vertical power}
\end{equation}
\noindent where $W = mg$ is the weight of UAV, with $m$ being its mass and $g$ representing the gravitational acceleration.

\par Accordingly, the total flight energy consumption of the $v$-th UAV in the $c$-th cluster is as follows:
\begin{equation}
E_{v_c}^{\text{flight}} = \int_{0}^{T} \left(P_{v_{c}}^{\textrm{xy}} + P_{v_{c}}^{\textrm{z}}\right) dt.
\label{eq. total flight energy consumption}
\end{equation}
\noindent where $T$ denotes the flight duration.

\section{Problem Formulation and Analysis}
\label{sec: Problem Formualtion and Analysis}

\noindent In this section, we formulate the optimization problem, termed data collection and semantic forwarding muti-objective optimization problem (DCSFMOP). Initially, the decision variables, objective functions, and overall structure of the optimization problem are presented. Then, the properties and characteristics of the formulated optimization problem are analyzed.

\subsection{Problem Formulation}
\label{subsec: Problem Formualtion}

\noindent The communication system considered in this work encompasses two distinct types of links, which are the ground user-to-UAV links and UAV cluster-to-remote BS links, respectively. In the first link type, UAVs must be spatially dispersed to conduct data collection from ground users. Conversely, the second link type necessitates a relatively compact spatial distribution among UAVs within a cluster, so that enabling the formation of a high-gain, directional main lobe aimed at the remote BS, thereby maximizing semantic communication performance. However, achieving these optimized aerial positions incurs considerable energy consumption from UAV flights. Therefore, how to achieve a balanced trade-off among three conflicting objectives, \emph{i.e.}, maximizing the volume of data collected from ground user, maximizing semantic communication performance between the UAV clusters and BS, and minimizing energy consumption of UAVs, is crucial for establishing an effective post-disaster communication network.

\par The decision variables are defined as follows. \emph{First}, the UAV clustering is denoted by $\mathbf{c} =$ $\{c_{v} \in \{1, 2, \ldots, N_{\mathrm{cluster}}\} $ $| v \in \mathcal{V}\}$ with $c_{v}$ indicating the cluster assignment of the $v$-th UAV. \emph{Second}, the set of UAV locations, composed of the continuous coordinates, is denoted by $\mathbf{Q} = \{\textbf{q}_{v_c} | v \in \mathcal{V}, c \in \mathcal{C}\}$. \emph{Third}, the set of excitation current weights of UAVs is denoted by $\mathbf{w} = \{w_{v, c} | v \in \mathcal{V}, c \in \mathcal{C}\}$. \emph{Finally}, the set of average number of semantic symbols used per word in each UAV cluster is denoted by $\mathbf{k} = \{k_{c, \mathrm{bs}} | c \in \mathcal{C}\}$. Based on these decision variables, the considered optimization objectives are detailed as follows.

\par In post-disaster communication scenarios, maximizing the total transmission rate of all served ground users is critical to ensure efficient and reliable data collection. Accordingly, the first optimization objective is formulated as
\begin{equation}
f_{1}(\mathbf{Q}) = \sum_{u=1}^{N_{U}} R_{u, v_u^\star}.
\label{eq. Optimization Objective 1}
\end{equation}

\par Considering the presence of multiple UAV clusters, the second optimization objective is to maximize the total semantic transmission rate, which is defined as the amount of semantic information sent per second and is measured in suts/s, thereby enhancing the overall network capacity. Thus, the second optimization objective function is as follows:
\begin{equation}
f_{2}(\mathbf{c}, \mathbf{Q}, \mathbf{w}, \mathbf{k}) = \sum_{c=1}^{N_{C}} SR_{c, \mathrm{bs}}.
\label{eq. Optimization Objective 2}
\end{equation}

\par Considering the limited battery capacity of UAVs and the impracticality of frequent recharging or replacement in post-disaster scenarios, the third optimization objective is to minimize the total energy consumption of all UAVs, which helps prolong the operational duration of the network and improves communication reliability. Therefore, the third objective function is defined as follows:
\begin{equation}
f_3(\mathbf{Q}) = \sum_{c=1}^{N_{\mathrm{cluster}}} \sum_{v=1}^{N_{V_c}} E_{v_{c}}^{\text{flight}}.
\label{eq. the third optimization objective}
\end{equation}

\par Accordingly, the DCSFMOP is formulated as follows:
\begin{subequations}
\label{eq. RPTRMOP}
\begin{align}
\underset{ \{\mathbf{c}, \mathbf{Q}, \mathbf{w}, \mathbf{k}\} }{\text{max}} \ & \{f_{1}, f_{2}, -f_{3}\}, \label{subeq. optimization problem} \\
\text{s.t.}\
& C1: \mathbf{q}_{v_c} \in \mathcal{R}, \quad \forall v \in \mathcal{V}, \forall c \in \mathcal{C}, \label{subeq. constraint 1}\\ 
& C2: \lVert \mathbf{q}_{{v_i}_{c_i}} - \mathbf{q}_{{v_j}_{c_j}} \rVert \geq d_{\textrm{min}}, \quad \forall v_i, v_j \in \mathcal{V}, v_i \neq v_j, \label{subeq. constraint 2}  \\
& C3: N_{\mathrm{V}_c} \ge 1, \quad \forall c \in \{1, \dots, N_{\textrm{cluster}}\}, \label{subeq. constraint 3} \\ 
& C4: c_{v} \in \{1, 2, \ldots, N_{\textrm{cluster}}\}, \quad \forall u \in \mathcal{U}, \label{subeq. constraint 4} \\
& C5: \sum_{c = 1}^{N_{\textrm{cluster}}} N_{\mathrm{V}_c} = N_{\textrm{V}}, \label{subeq. constraint 5}\\
& C6: \xi_{c, \mathrm{bs}} \geq \xi_{\text{th}}, \forall c \in \mathcal{C}, \label{subeq. constraint 6} \\
& C7: k_{c,s} = 
\begin{cases}
[k_{\min}, k_{\max}] \quad \text{ if } f_1 > f_2 \\
\textrm{not optimized} \quad \text{ if } f_1 \le f_2,
\end{cases} \label{subeq. constraint 7}
\end{align}
\end{subequations}
\noindent where constraint \eqref{subeq. constraint 1} restricts the deployment of UAVs to the predefined 3D space $\mathcal{R}$ and constraint \eqref{subeq. constraint 2} manifests a minimum safety distance between any two UAVs to avoid collisions. Moreover, constraint \eqref{subeq. constraint 3} enforces that each cluster contains at least one UAV, constraint \eqref{subeq. constraint 4} guarantees that each UAV belongs to exactly one cluster, constraint \eqref{subeq. constraint 5} ensures that the total number of UAVs across all clusters equals the total number of available UAVs, and constraint \eqref{subeq. constraint 6} defines the minimum semantic rate threshold $\xi_{\text{th}}$. In addition, constraint \eqref{subeq. constraint 7} indicates that when the total transmission rate of all served ground users $f_{1}$ exceeds the total semantic transmission rate $f_{2}$, $k_{c, \mathrm{bs}}$ is optimized within the range $[k_{\mathrm{min}}, k_{\mathrm{max}}]$. Conversely, when $f_{1} \leq f_{2}$, $k_{c, \mathrm{bs}}$ will not be optimized since semantic processing can evidently be omitted.

\subsection{Problem Analysis}
\label{subsec: Problem Analysis}

\noindent In this section, the formulated DCSFMOP is analyzed as follows.

\par \emph{(i) The formulated optimization problem in Eq. \eqref{subeq. optimization problem} is an MINLP problem.} \underline{\emph{First}}, the involvement of both discrete variables (clustering assignment $\mathbf{c}$ and semantic symbol number $\mathbf{k}$) and continuous variables (UAV location $\mathbf{Q}$ and excitation current weight $\mathbf{w}$) classifies it a mixed-integer problem. \underline{\emph{Second}}, the objective functions $f_1$, $f_2$, and $f_3$ are all inherently nonlinear. Specifically, $f_1$ and $f_2$ apply nonlinear distance-based path loss model and $f_3$ contains square roots, power functions, and integrals which are nonlinear. Therefore, the formulated DCSFMOP is a typical MINLP problem.

\par \emph{(ii) The formulated optimization problem in Eq. \eqref{subeq. optimization problem} is NP-hard.} Specifically, if $\mathbf{Q}, \mathbf{w}, \mathbf{k}$ are fixed, the optimization with respect to the discrete cluster assignment $\mathbf{c}$ reduces to allocating $N_V$ UAVs to $N_{\mathrm{cluster}}$ clusters, subject to the constraints that each UAV must belong to exactly one cluster and each cluster must contain at least one UAV, while aiming to maximize $f_1$ and $f_2$. In this case, $f_2$ imposes a requirement that the UAVs within each cluster maintain a compact spatial configuration to maximize the array gain. The aforementioned subproblem can be formulated as a k-means clustering problem, which is known to be NP-hard. Therefore, the subproblem is NP-hard, and consequently, the overall DCSFMOP is also NP-hard.

\par \emph{(iii) The formulated optimization problem in Eq. \eqref{subeq. optimization problem} has dynamically varying dimensionality of decision variables.} Specifically, the optimization of the number of semantic symbols per cluster, \emph{i.e.}, $\mathbf{k}$, is determined by the UAV cluster assignment scheme, as the dimensionality of $\mathbf{k}$ can only be determined once the number of clusters is specified. Consequently, any changes in the number of UAV clusters results in a corresponding change in the dimension of $\mathbf{k}$, which in turn leads to dynamically varying decision variables throughout the optimization process.

\par Given the aforementioned characteristics, the formulated optimization problem cannot be directly tackled by existing conventional algorithms. Therefore, we propose a dedicated approach to efficiently obtain feasible solutions.

%
%
\section{The Proposed Approach}
\label{sec: Algorithm}

\noindent In this section, we first elaborate the motivation of proposing LLM-AOA for solving the formulated problem. Then, we present the preliminaries of the proposed LLM-AOA. Finally, we introduce the detailed description of the overall LLM-AOA framework.

\subsection{Motivation}
\label{subsec: Motivation}

\noindent As presented in Section \ref{subsec: Problem Analysis}, the formulated DCSFMOP is a MINLP problem and inherently NP-hard with dynamically varying decision variable dimensionality, which renders it intractable for traditional algorithms. Specifically, DRL-based algorithms face extremely difficult state representation and action space definition when dealing with dynamic decision spaces, while convex optimization-based algorithms are generally unsuitable since the formulated problem is non-convex. In this case, AO provides an effective framework for the formulated DCSFMOP by sequentially optimizing different subsets of decision variables. First, it determines the UAV clustering assignment, which fixes the dimensionality of the remaining decision variables. Then, the remaining variables are optimized sequentially. The approach mitigates the computational complexity of jointly handling dynamically varying variables.

\par Different subsets of decision variables exhibit distinct mathematical properties, which necessitates tailored optimization strategies. Specifically, the UAV clustering assignment is a combinatorial problem with an exponentially growing search space, for which a greedy strategy offers an optimal trade-off between computational efficiency and solution quality. Moreover, for optimizing the number of semantic symbols, which is a low-dimensional discrete problem, a greedy strategy yields optimal or near-optimal solutions with relatively low computational overhead. Finally, evolutionary computation algorithms are suitable for the joint optimization of UAV locations and weights which is a high-dimensional, continuous, and non-convex problem~\cite{Sun2021Time}. However, these algorithms often struggle with premature convergence, parameter sensitivity, and inefficient exploration in complex solution spaces.

\par The integration of LLMs into the optimization process of evolutionary computation algorithms addresses these limitations through their advanced reasoning and analytical capabilities. LLMs offer the following advantages in our research context. \emph{First}, LLMs dynamically analyze population statistics and evolutionary patterns to identify optimization states that remain undetectable through conventional metrics. \emph{Second}, they also leverage transfer learning from diverse domains to propose effective parameter adjustments appropriate for different optimization phases. \emph{Finally}, LLMs provide context-aware guidance that adapts to the evolving search landscape. When algorithms exhibit premature convergence, LLMs recommend increasing the value of mutation parameter or implementing diversity preservation mechanisms. Conversely, in situations of slow convergence with excessive diversity, LLMs suggest decreasing the value of mutation parameter. Such a dynamic parameter adjustment mechanism enhances algorithmic navigation through the complex search space of UAV locations and weights.

\par According to the aforementioned motivation, we propose LLM-AOA, which synergistically integrates greedy strategy and LLM-guided evolutionary computation within the AO framework to efficiently and effectively solve the formulated DCSFMOP. In the following, some preliminaries of the proposed LLM-AOA are introduced.

\subsection{Preliminaries}
\label{subsec: Preliminaries}

\noindent In this section, we present the principles of the AO framework, greedy strategy, evolutionary computation, and large language model, respectively.

\subsubsection{AO Framework}
\label{subsubsec: Alternating Optimization}

\noindent AO decomposes a complex optimization problem into several subproblems by iteratively optimizing one subset of decision variables while keeping others fixed. Given an optimization problem $f(x_1, \ldots, x_n)$ with decision variables $x_1, \ldots, x_n$, where $n$ is the dimension of the decision variables, the AO procedure proceeds as follows~\cite{bezdek2003convergence}:

\par \textbf{1) Initialization.} Initialize variables $\{x_i^{(0)}\}_{i=1}^{n}$ and set the maximum number of iterations $T_{\mathrm{max}}^{\mathrm{AO}}$.

\par \textbf{2) Alternating Optimization.} In each iteration, the variables are optimized sequentially while all other variables remain fixed. Specifically, in the $t$-th iteration, $x_1^{(t)}$ is updated as follows:
\begin{equation}
x_1^{(t)} = \arg\min_{x_1} f(x_1, x_2^{(t-1)}, \ldots, x_n^{(t-1)}).
\label{eq. alternating optimization update x_1}
\end{equation}

\par Subsequently, $x_2^{(t)}$ is optimized as follows:
\begin{equation}
x_2^{(t)} = \arg\min_{x_2} f(x_1^{(t)}, x_2, \ldots, x_n^{(t-1)}).
\label{eq. alternating optimization update x_1}
\end{equation}

\par The process continues until all decision variables are updated within the iteration.

\par \textbf{3) Termination.} The algorithm terminates when the maximum number of iterations $T_{\mathrm{max}}^{\mathrm{AO}}$ is reached, yielding the final variables $\{x_i^{(T_{\mathrm{max}}^{\mathrm{AO}})}\}_{i=1}^{n}$. Alternatively, convergence-based stopping criteria can also be used, such as $\|x_i^{(t)} - x_i^{(t-1)}\| \leq \epsilon_1$ for all $i \in \{1, 2, \ldots, n\}$.

\subsubsection{Greedy Strategy}
\label{subsubsec: Greedy Strategy}

\noindent The greedy strategy selects the locally optimal choice at each step based on the currently available information. The fundamental principle lies in constructing a global solution through a sequence of locally optimal decisions, under the assumption that such incremental optimization will lead to a near-optimal or optimal overall solution. Specifically, the greedy strategy is characterized by the following two properties:

\par \textbf{1) Greedy Choice.} Given an original problem $\mathbb{P}$, let $x_1$ be the greedy choice made at the current step. Let $\mathbb{P}'$ denote the resulting subproblem after this choice is incorporated. The greedy-choice property guarantees that there exists an optimal solution $\mathbf{A}$ to $\mathbb{P}$ that can be expressed as $\mathbf{A} = \{x_1, \mathbf{A}'\}$, where $\mathbf{A}'$ is an optimal solution to the subproblem $\mathbb{P}'$.

\par \textbf{2) Optimal Substructure.} Let $\mathbb{P}$ be the original optimization problem and $\mathbf{A}$ be an optimal solution to $\mathbb{P}$. If $\mathbf{A}$ is composed of the current choice $x$ and a solution $\mathbf{A}'$ of the subproblem $\mathbb{P}'$, then $\mathbf{A}'$ must be the optimal solution of the subproblem $\mathbb{P}'$.

\subsubsection{Evolutionary Computation}
\label{subsubsec: Evolutionary Computation}

\noindent  Evolutionary computation is a class of optimization methods inspired by natural evolution and swarm intelligence. By iteratively updating a population of candidate solutions, the overall performance gradually improves~\cite{back2002evolutionary}. Evolutionary computation is suitable for high-dimensional, non-convex, multi-objective, and complex constraint optimization problems.

\par The procedure of evolutionary computation algorithms is as follows. First, an initial population of candidate solutions is generated. Then, the algorithm iteratively produces new candidates, evaluates their fitness, and selects superior solutions to form the next generation. The iterative process continues until a predefined termination condition, e.g., a maximum number of iterations is satisfied, upon which the algorithm outputs an optimal or near-optimal solution.

\subsubsection{Large Language Model}
\label{subsubsec: LLM}

\noindent LLMs are a class of foundation models trained on large-scale corpora, which are capable of capturing complex linguistic structures, contextual semantic relationships, and implicit high-level reasoning patterns~\cite{Zheng2025LLM}. Representative LLMs include the GPT series~\cite{achiam2023gpt}, e.g., GPT-3.5 and GPT-4, Gemini~\cite{team2023gemini}, and Claude, developed by OpenAI, Google, and Anthropic, respectively. 

\par The capability of LLMs to comprehend, summarize, and support decision-making on complex data enables them to assist in addressing diverse optimization problems. Specifically, LLMs can act as intelligent coordinators for optimization algorithms, which dynamically adjusts parameters and control strategies to guide the algorithms towards more efficient and effective solutions~\cite{LLM2025optimization},~\cite{wang2025llm}. For example, they can adjust reward function weights in DRL-based algorithms and tune population update rules in evolutionary computation.

\subsection{The Proposed LLM-AOA}
\label{subsec: The Proposed LLM-AOA}

\noindent Based on the aforementioned preliminaries, we propose LLM-AOA to effectively solve the formulated DCSFMOP. Inspired by evolutionary computation, LLM-AOA adopts a population-based solution representation, where each individual denotes a feasible solution. Specifically, LLM-AOA comprises an initialization phase which generates the initial population and an AO phase which evolves the population to obtain high-quality solutions.

\par Fig. \ref{fig. The framework of the proposed LLM-AOA} shows the framework of the proposed LLM-AOA in terms of handling the formulated DCSFMOP, which provides a visual representation of how these components are integrated. The overall structure of LLM-AOA is shown in Algorithm~\ref{algo: The proposed LLM-AOA} and the details are elaborated as follows.

\begin{figure*}
\centerline{\includegraphics[width=6.9in]{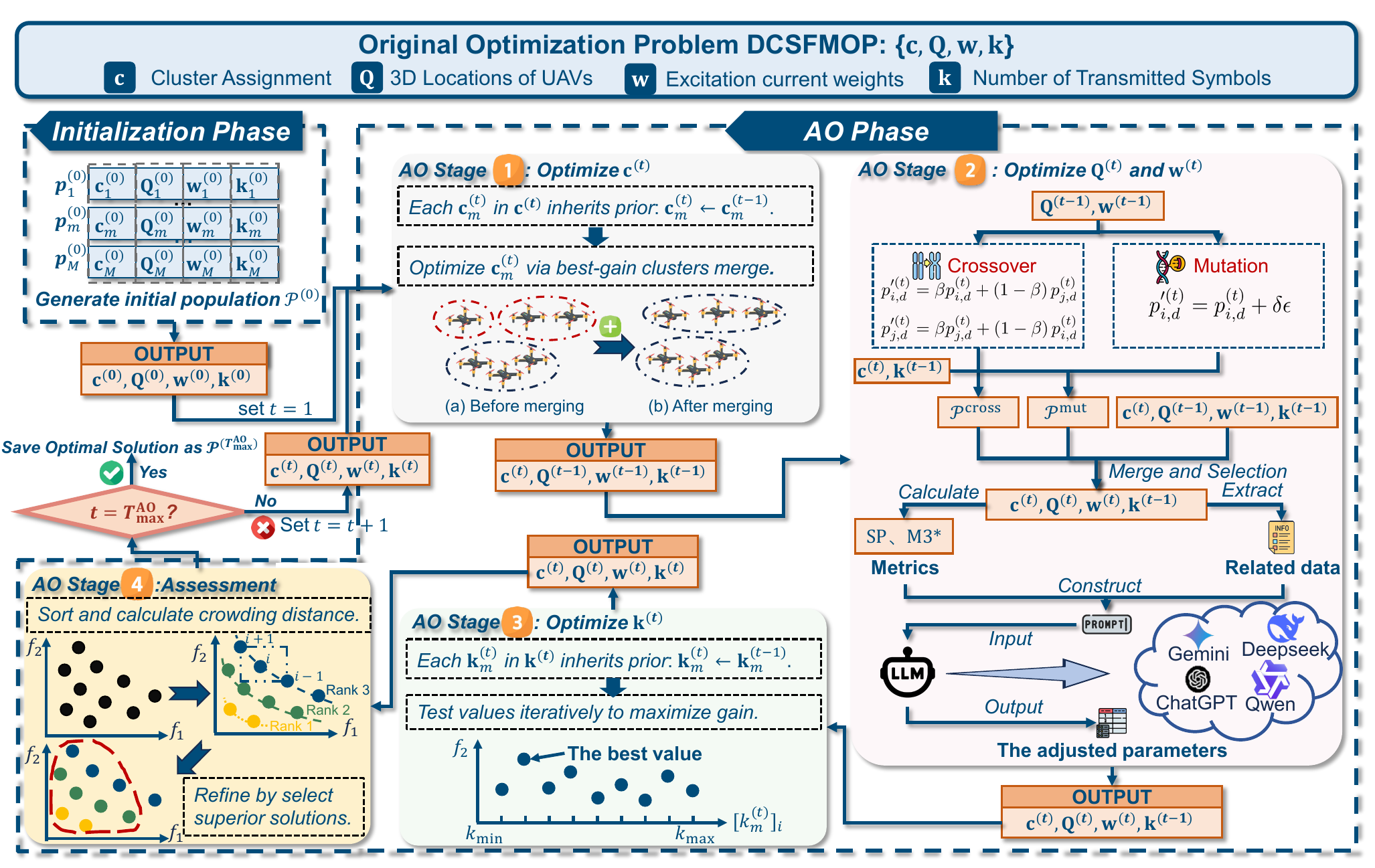}}
\caption{The framework of the proposed LLM-AOA. In the initialization phase, the initial population $\mathcal{P}^{(0)}$ is generated. Then, in each iteration of the AO phase, the UAV clustering assignment $\mathbf{c}$, UAV locations $\mathbf{Q}$ and excitation current weights $\mathbf{w}$, and the number of semantic symbols per UAV cluster $\mathbf{k}$ are sequentially optimized.}
\label{fig. The framework of the proposed LLM-AOA}
\end{figure*}

\subsubsection{Initialization Phase}
\label{subsubsec: Initialization Phase}

\noindent During the initialization phase, LLM-AOA generates an initial population $\mathcal{P}^{(0)}$ and sets the algorithmic parameters. Given the heterogeneous decision variables, the traditional unified initialization is infeasible. Therefore, LLM-AOA adopts a mixed initialization method to cover the diverse search space and enhance the uniformity of solution distribution. Specifically, the population size is set as $M$ and then the $m$-th individual in $\mathcal{P}^{(0)}$ is as follows:
\begin{equation}
\label{eq. LLM-AOA initialization}
\boldsymbol{p}_m^{(0)} = \{\mathbf{c}_m^{(0)}, \mathbf{Q}_m^{(0)}, \mathbf{w}_m^{(0)}, \mathbf{k}_m^{(0)}\}, \quad m=1,\ldots, M.
\end{equation}

\par Given the heterogeneous nature of the decision variables, \emph{i.e.}, $\mathbf{c}_m^{(0)}$, $\mathbf{Q}_m^{(0)}$, $\mathbf{w}_m^{(0)}$, $\mathbf{k}_m^{(0)}$, we initialize each component separately.

\par \textbf{1) Initialization of $\mathbf{c}_m^{(0)}$:} The $i$-th element $[c_m^{(0)}]_i$ of the UAV cluster assignment vector $\mathbf{c}_m^{(0)}$ is initialized as follows:
\begin{equation}
\label{eq. UAV cluster assignment initialization}
[c_m^{(0)}]_i = \mathrm{randint}(1, N_{\mathrm{V}}),
\end{equation}
\noindent where $\mathrm{randint}(1, N_{\mathrm{V}})$ denotes the function that selects an integer uniformly at random from the discrete set $\{1, \ldots, N_{\mathrm{V}}\}$. 

\par \textbf{2) Initialization of $\mathbf{Q}_m^{(0)}$:} The 3D locations of the $i$-th UAV $[\mathbf{q}^{(0)}_{m}]_{i}$ in $\mathbf{Q}_m^{(0)} = \{[\mathbf{q}^{(0)}_{m}]_{1}, \ldots, [\mathbf{q}^{(0)}_{m}]_{N_{\mathrm{V}}}\}$ is initialized as follows:
\begin{equation}
[\mathbf{q}^{(0)}_{m}]_{i} = \mathbf{q}_{\mathrm{min}} + \mathbf{r} \odot (\mathbf{q}_{\mathrm{max}} - \mathbf{q}_\mathrm{min}),
\end{equation}
\noindent where $\mathbf{r}$ represents a 1×3 array whose elements are independently drawn from a uniform distribution over the interval $[0, 1]$ and $\odot$ is the Hadamard (element-wise) product. The vector $\mathbf{q}_{\mathrm{min}} = \{x_{\min}, y_{\min}, z_{\min}\}$ and $\mathbf{q}_{\mathrm{max}} = \{x_{\max}, y_{\max}, z_{\max}\}$ define the lower and upper bounds of the 3D locations for all UAVs, respectively.

\begin{algorithm}[htbp]

    \caption{The proposed LLM-AOA}
    \label{algo: The proposed LLM-AOA}
    
    \KwIn{$N_{\textrm{V}}$, $N_{\textrm{U}}$, $M$, $T_{\mathrm{max}}^{\mathrm{AO}}$.}  
    \KwOut{$\mathcal{P}^{(T_{\mathrm{max}}^{\mathrm{AO}})}$.}
  
    \tcc{Initialization Phase}  
    Set $\mathcal{P}^{(0)} \gets \varnothing$;   \\
    \For{$m = 1$ \textbf{to} $M$}
    {
        Initialize $\boldsymbol{p}_m^{(0)} = \{\mathbf{c}_m^{(0)}, \mathbf{Q}_m^{(0)}, \mathbf{w}_m^{(0)}, \mathbf{k}_m^{(0)}\}$;  \\
        
        Set $\mathcal{P}^{(0)} \gets \mathcal{P}^{(0)} \bigcup \boldsymbol{p}_m^{(0)}$;   \\
    }  
    \tcc{AO Phase}
    \For{$t = 1$ \textbf{to} $T_{\mathrm{max}}^{\mathrm{AO}}$} 
    {   
        \tcc{Stage 1: Optimize $\mathbf{c}^{(t)}$}
        $\mathbf{c}^{(t)} \gets$ GCA($\mathbf{c}^{(t-1)}$, $\mathbf{Q}^{(t-1)}$,
        $\mathbf{w}^{(t-1)}$, $\mathbf{k}^{(t-1)}$);
    
        \tcc{Stage 2: Optimize $\mathbf{Q}^{(t)}$ and $\mathbf{w}^{(t)}$}
        $\mathbf{Q}^{(t)}$, $\mathbf{w}^{(t)}$ $\gets$ LLM-guided NSGA-II($\mathbf{c}^{(t)}$, $\mathbf{Q}^{(t-1)}$, $\mathbf{w}^{(t-1)}$, $\mathbf{k}^{(t-1)}$); 
 
        \tcc{Stage 3: Optimize $\mathbf{k}^{(t)}$}
        $\mathbf{k}^{(t)}$ $\gets$ GSO($\mathbf{c}^{(t)}$, $\mathbf{Q}^{(t)}$, $\mathbf{w}^{(t)}$, $\mathbf{k}^{(t-1)}$); 
        
        \tcc{Stage 4: Assessment}
        Conduct non-dominated sorting on all individuals in $\mathcal{P}^{(t)} = \{\boldsymbol{p}_{1}^{(t)}, \ldots,  \boldsymbol{p}_{m}^{(t)}, .., \boldsymbol{p}_{M}^{(t)}\}$; \\
        
        Calculate crowding distance of individuals in $\mathcal{P}^{(t)}$;\\
        
        Sort $\mathcal{P}^{(t)}$ and truncate redundant individual in $\mathcal{P}^{(t)}$;\\
        
    }
    Return $\mathcal{P}^{(T_{\mathrm{max}}^{\mathrm{AO}})}$.
    
\end{algorithm}

\begin{algorithm}[htbp]
\caption{GCA}
\label{algo: greedy_clustering}

\KwIn{$\mathbf{c}^{(t-1)}$, $\mathbf{Q}^{(t-1)}$, $\mathbf{w}^{(t-1)}$, $\mathbf{k}^{(t-1)}$, $M$. } 
\KwOut{$\mathbf{c}^{(t)}$.}

\For{$m = 1$ \textbf{to} $M$}
{
    \tcc{Initialization Process}
    Set $\mathbf{c}^{(t)}_{m} \gets \mathbf{c}^{(t-1)}_{m}$;\\
    
    Calculate $N^{(t)}_{\mathrm{cluster}}$ of $\mathbf{c}^{(t)}_{m}$ via Eq.~\eqref{eq. calculation of the number of UAV clusters}; \\ 
    
    Initialize the set of UAV clusters in the $t$-th iteration $\mathscr{C}^{(t)} = \{\mathcal{C}^{(t)}_1, \mathcal{C}^{(t)}_2, \ldots, \mathcal{C}^{(t)}_{N^{(t)}_{\mathrm{cluster}}} \}$;\\
    
    \For{$i = 1$ \textbf{to} $N^{(t)}_{\mathrm{cluster}}$}
    {
        $\mathcal{C}^{(t)}_i = \left\{ j \mid c^{(t)}_j = i, \ 1 \leq j \leq N_{\text{V}} \right\}$;
    } 
    
    Set $\texttt{continue} \gets \texttt{True}$; \\     
    
    \tcc{Greedy Iterative Process}
    \While{$\texttt{continue}$}
    {
    
    Set $f_2^{\mathrm{gain}} = -\infty$;\\
    
    Set $\mathbf{c}^{\mathrm{merge}} = \varnothing$;
    
    \For{$b, b' \in \{1,\ldots, |\mathscr{C}^{(t)}|\},\, b \ne b'$}
    {   
        
        Create a copy $\mathscr{C}^{\mathrm{temp}} = \mathscr{C}^{(t)}$;\\
    
        Merge clusters as $\mathcal{C}^{\mathrm{temp}}_{b} = \mathcal{C}^{\mathrm{temp}}_{b} \bigcup \mathcal{C}^{\mathrm{temp}}_{b'}$; \\
        
        Remove $\mathcal{C}^{\mathrm{temp}}_{b'}$ from $\mathscr{C}^{\mathrm{temp}}$;\\
    
        \For{$i = 1$ \textbf{to} $|\mathscr{C}^{\mathrm{temp}}|$}
        {
            Re-number the cluster index in $\mathscr{C}^{\mathrm{temp}}$;
        }
        
        Update $\mathbf{c}^{\mathrm{temp}}$ based on $\mathscr{C}^{\mathrm{temp}}$;\\
    
        Calculate $f_2^\mathrm{temp}$ based on the updated $\mathbf{c}^{\mathrm{temp}}$, $\mathbf{Q}^{(t-1)}$, $\mathbf{w}^{(t-1)}$, and $\mathbf{k}^{(t-1)}$;\\
        
        \If{$f_2^\mathrm{temp} - [{f_2}]_{m}^{(t-1)} > f^{\mathrm{gain}}_2 $}
        {   
            $\mathbf{c}^{\mathrm{merge}} = \{b, b'\}$ ; \\
            Set $f^{\mathrm{gain}}_2 = f_2^\mathrm{temp} - [{f_2}]_{m}^{(t-1)}$; \\
        }     
        
    }    
    \If{$f^{\mathrm{gain}}_2 > 0$}   
    {
        Extract indices $(b, b')$ from $\mathbf{c}^{\mathrm{merge}}$;\\
    
        Merge clusters: $\mathcal{C}^{(t)}_{b} \gets \mathcal{C}^{(t)}_{b} \cup \mathcal{C}^{(t)}_{b'}$;\\
    
        Remove $\mathcal{C}^{(t)}_{b'}$ from $\mathscr{C}^{(t)}$;\\

        \For{$i = 1$ \textbf{to} $|\mathscr{C}^{(t)}|$}
        {
            Re-number the cluster index in $\mathscr{C}^{(t)}$;
        }
        
        Update $\mathbf{c}^{(t)}_{m}$ based on $\mathscr{C}^{(t)}$;\\
    
    } 
    \Else {
        Set $\texttt{continue} = \texttt{False}$; \\    
    } 

}
}
Return $\mathbf{c}^{(t)}$.
\end{algorithm}

\par \textbf{3) Initialization of $\mathbf{w}_m^{(0)}$:} The $i$-th element $[w_m^{(0)}]_i$ of the excitation current weight vector $\mathbf{w}_m^{(0)}$ is initialized as follows:
\begin{equation}
\label{eq. UAV weight initialization}
[w_m^{(0)}]_i = \mathrm{uniform}(w_{\min}, w_{\max}),
\end{equation}
\noindent where $\mathrm{uniform}(w_{\min}, w_{\max})$ denotes a function that selects a real number from the continuous interval $[w_{\min}, w_{\max}]$ based on the uniform probability distribution.

\par \textbf{4) Initialization of $\mathbf{k}_m^{(0)}$:} The $i$-th element $[k_m^{(0)}]_i$ of the number of transmitted symbols of each cluster $\mathbf{k}_m^{(0)}$ is initialized by selecting a discrete integer uniformly at random from the range $[k_{\min}, k_{\max}]$, the initialization is as follows:
\begin{equation}
\label{eq. cluster K initialization}
[k_m^{(0)}]_i = \mathrm{randint}(k_{\min}, k_{\max}).
\end{equation}

\par After all the aforementioned components, \emph{i.e.}, $\mathbf{c}_m^{(0)}$, $\mathbf{Q}_m^{(0)}$, $\mathbf{w}_m^{(0)}$, and $\mathbf{k}_m^{(0)}$, are initialized, they are concatenated to form the $m$-th individual $\boldsymbol{p}_m^{(0)}$. The concatenation of all individuals, i.e., $\boldsymbol{p}_1^{(0)}, \ldots, \boldsymbol{p}_M^{(0)}$, forms the initial population $\mathcal{P}^{(0)}$. Consequently, an alternative representation of $\mathcal{P}^{(0)}$ can be formulated as $\mathcal{P}^{(0)} = \{\mathbf{c}^{(0)}, \mathbf{Q}^{(0)}, \mathbf{w}^{(0)}, \mathbf{k}^{(0)} \}$, where $\mathbf{c}^{(0)} = \{\mathbf{c}_1^{(0)}, \ldots, \mathbf{c}_M^{(0)}\}$, $\mathbf{Q}^{(0)} = \{\mathbf{Q}_1^{(0)}, \ldots, \mathbf{Q}_M^{(0)}\}$, $\mathbf{w}^{(0)} = \{\mathbf{w}_1^{(0)}, \ldots, \mathbf{w}_M^{(0)}\}$, and $\mathbf{k}^{(0)} = \{\mathbf{k}_1^{(0)}, \ldots, \mathbf{k}_M^{(0)}\}$.

\par Finally, $\mathcal{P}^{(0)} = \{\mathbf{c}^{(0)}, \mathbf{Q}^{(0)}, \mathbf{w}^{(0)}, \mathbf{k}^{(0)} \}$ is input to the AO phase for further optimization.

\subsubsection{AO Phase}
\label{subsubsec: Optimization Phase}

\noindent During the optimization phase, we use the AO framework to address the varying decision variables dimensionality. Specifically, the decision variables are decomposed into three components, \emph{i.e.}, $\mathbf{c}^{(t)}$, $\mathbf{Q}^{(t)}$ as well as $\mathbf{w}^{(t)}$, and $\mathbf{k}^{(t)}$. Accordingly, the overall AO process is systematically structured into four sequential stages. \emph{In AO stage 1, $\mathbf{c}^{(t)}$ is optimized by a greedy strategy-based algorithm. In AO stage 2, $\mathbf{Q}^{(t)}$ and $\mathbf{w}^{(t)}$ are optimized via an evolutionary computation algorithm which is assisted by LLM. In AO stage 3, $\mathbf{k}^{(t)}$ is determined by using another greedy strategy-based algorithm. Finally, in AO stage 4, comprehensive assessment is conducted to filter superior solutions.} The details of the AO phase are presented as follows.

\par \textbf{1) AO Stage 1: Optimization of $\mathbf{c}^{(t)}$:} The optimization of $\mathbf{c}^{(t)}$ is a combinatorial optimization problem whose solution space grows exponentially with the number of UAVs increasing. Compared with exhaustive search or conventional evolutionary algorithms, a greedy strategy achieves high-quality solutions with significantly lower computational overhead. Therefore, in this stage, $\mathbf{c}^{(t)}$ is optimized by the proposed GCA, while the other decision variables, \emph{i.e.}, $\mathbf{Q}^{(t-1)}$, $\mathbf{w}^{(t-1)}$, and $\mathbf{k}^{(t-1)}$, are fixed. \emph{Specifically, the proposed GCA iteratively optimizes the UAV cluster assignment of each individual in the population, i.e., $\mathbf{c}_1^{(t)}, \ldots, \mathbf{c}_M^{(t)}$.} The algorithm is presented in Algorithm \ref{algo: greedy_clustering} and the involved initialization and greedy iterative processes are elaborated as follows.

\par \emph{\textbf{During the initialization process}}, the proposed GCA initializes the cluster assignment, constructs the set of clusters, and declares the control variables for the forthcoming greedy iterative process.

\par \emph{First}, the proposed GCA initializes the cluster assignment of the current iteration as $\mathbf{c}_{m}^{(t)} \gets \mathbf{c}_{m}^{(t-1)}$, where $\mathbf{c}_{m}^{(t)} = \{c_1^{(t)}, c^{(t)}_2, \ldots, c_{N_{\mathrm{V}}}^{(t)} \}$. 

\par \emph{Subsequently}, the total number of UAV clusters in $\mathbf{c}_{m}^{(t)}$ is determined by the maximum cluster index as follows:
\begin{equation}
\label{eq. calculation of the number of UAV clusters}
N^{(t)}_{\mathrm{cluster}} = \mathrm{max}\left(c^{(t)}_1, c^{(t)}_2, \ldots, c^{(t)}_{N_{\mathrm{V}}} \right),
\end{equation}
\noindent where $\mathrm{max}(\cdot)$ returns the maximum value among the elements in the input set. Based on the calculated $N^{(t)}_{\mathrm{cluster}}$, the set of UAV clusters in the $t$-th iteration, denoted by $\mathscr{C}^{(t)}$, is initialized as $\mathscr{C}^{(t)} = \{\mathcal{C}^{(t)}_1, \mathcal{C}^{(t)}_2, \ldots, \mathcal{C}^{(t)}_{N^{(t)}_{\mathrm{cluster}}} \}$. Initially, each element in $\mathscr{C}^{t}$ is set to the empty set $\varnothing$. Then, each element of $\mathscr{C}^{(t)}$ is initialized. For example, the set of $\mathcal{C}^{(t)}_i$, which contains the indices of all UAVs belonging to the $i$-th cluster, is set as $\mathcal{C}^{(t)}_i = \left\{ j \mid c^{(t)}_j = i, \ 1 \leq j \leq N_{\text{V}} \right\}$;

\par \emph{Finally}, the proposed GCA initializes a boolean variable $\texttt{continue} = \texttt{TRUE}$ to regulate the iterative execution of the subsequent greedy process.

\par \emph{\textbf{During the greedy iterative process}}, the proposed GCA searches over all cluster pairs to identify the merge operation which yields the greatest enhancement in the objective function $f_2$.

\par \emph{First}, several auxiliary parameters are declared. Specifically, the benefit gain $f^{\mathrm{gain}}_2$ of $f_2$ in the current loop is initialized to $-\infty$ and the set of cluster pairs yielding the maximum merging gain is defined as $\mathbf{c}_{\mathrm{merge}}$. 

\par \emph{Subsequently}, clusters are iteratively merged in a pairwise manner. Specifically, for each candidate pair of clusters indexed by $b$ and $b'$, a temporary cluster set $\mathscr{C}^{\mathrm{temp}}$ is constructed based on the current $\mathscr{C}^{(t)}$. The merging operation incorporates the $b'$-th cluster $\mathcal{C}^{\mathrm{temp}}_{b'}$ into the $b$-th cluster $\mathcal{C}^{\mathrm{temp}}_{b}$ via $\mathcal{C}^{\mathrm{temp}}_{b} \gets \mathcal{C}^{\mathrm{temp}}_{b} \bigcup \mathcal{C}^{\mathrm{temp}}_{b'}$, after which the redundant $b'$-th cluster $\mathcal{C}^{\mathrm{temp}}_{b'}$ is removed from $\mathscr{C}^{\mathrm{temp}}$ by $\mathscr{C}^{\mathrm{temp}} = \mathscr{C}^{\mathrm{temp}} \setminus \mathcal{C}^{\mathrm{temp}}_{b'}$. In the following, the cluster labels are then readjusted to ensure that all indices are consecutive integers. \emph{For example}, assuming $\mathscr{C}^{\mathrm{temp}} = \{\mathcal{C}^{\mathrm{temp}}_1, \mathcal{C}^{\mathrm{temp}}_2, \mathcal{C}^{\mathrm{temp}}_3\}$, in this case, if $\mathcal{C}^{\mathrm{temp}}_1$ and $\mathcal{C}^{\mathrm{temp}}_2$ are merged, the elements in $\mathscr{C}^{\mathrm{temp}}$ will be adjusted as $\mathscr{C}^{\mathrm{temp}} = \{\mathcal{C}^{\mathrm{temp}}_1, \mathcal{C}^{\mathrm{temp}}_2\}$. Following the reorganization, $\mathbf{c}^{\mathrm{temp}}$ is updated based on $\mathscr{C}^{\mathrm{temp}}$, and the corresponding objective function value $f^{\mathrm{temp}}_2$ is calculated. The algorithm then evaluates whether $f^{\mathrm{temp}}_2 - [{f_2}]_{m}^{(t-1)} $ exceeds $f^{\mathrm{gain}}_2$. If this condition is satisfied, indicating that merging the $b$-th and $b'$-th clusters provides a superior benefit, the algorithm updates $\mathbf{c}^{\mathrm{merge}} = \{b, b'\}$ and revises the gain metric as follows:
\begin{equation}
f^{\mathrm{gain}}_2 = f_2^\mathrm{temp} - [{f_2}]_{m}^{(t-1)}.
\end{equation}

\par \emph{Finally}, following the evaluation of all potential cluster pairs, the GCA determines whether to perform a merge operation based on the auxiliary parameter $f^{\mathrm{gain}}_2$. If $f^{\mathrm{gain}}_2 > 0$, indicating that a beneficial merging exists, the algorithm proceeds with the optimal merger in $\mathbf{c}^{\mathrm{merge}}$. Specifically, the algorithm extracts the corresponding cluster indices $b$ and $b'$ from $\mathbf{c}^{\mathrm{merge}}$, incorporates the $b'$-th cluster into the $b$-th cluster, removes the $b'$-th cluster from $\mathscr{C}^{(t)}$, renumbers the remaining clusters in $\mathscr{C}^{(t)}$, and updates $\mathbf{c}^{(t)}$ based on the merged cluster set $\mathscr{C}^{(t)}$. Conversely, if $f^{\mathrm{gain}}_2 \leq 0$, the Boolean variable \texttt{continue} is set to \texttt{false}, and the GCA terminates. 

\par Based on the aforementioned GCA, $\mathbf{c}^{(t)}$ is optimized. The optimized $\mathbf{c}^{(t)}$, along with $\mathbf{Q}^{(t-1)}$, $\mathbf{w}^{(t-1)}$, and $\mathbf{k}^{(t-1)}$ from the previous iteration, are then input to \textbf{AO Stage 2} for the optimization of $\mathbf{Q}^{(t)}$ and $\mathbf{w}^{(t)}$ in current iteration.

\begin{algorithm}[htbp]
\caption{LLM-guided NSGA-II}
\label{algo: NSGA-II}

\KwIn{$\mathbf{c}^{(t)}, \mathbf{Q}^{(t-1)}, \mathbf{w}^{(t-1)}, \mathbf{k}^{(t-1)}, M$.}
\KwOut{$\mathbf{Q}^{(t)}$, $\mathbf{w}^{(t)}$.}

\tcc{Initialization Process}
Set parameters $T_{\mathrm{local}}$, $p_{\mathrm{c}}$, and $p_{\mathrm{m}}$;\\
Set $\mathcal{P}^{\mathrm{temp}} \gets \{\mathbf{c}^{(t)}, \mathbf{Q}^{(t-1)}, \mathbf{w}^{(t-1)}, \mathbf{k}^{(t-1)} \}$;\\

\tcc{Evolution Process}
\For{$t = 1$ \KwTo $T_{\mathrm{local}}$}{

    \tcc{Crossover Operation}
    Set $n_{\mathrm{cross}} = p_{\mathrm{c}}M$; \\
    Initialize the crossover population $\mathcal{P}^{\mathrm{cross}} \gets \varnothing$; \\
    \For{$k = 1$ \KwTo $n_{\mathrm{cross}}/2$}{
        Select parents $\boldsymbol{p}_1$, $\boldsymbol{p}_2$ from $\mathcal{P}^{\mathrm{temp}}$;\\
        Apply crossover on $(\mathbf{Q}^{(t-1)},\mathbf{w}^{(t-1)})$ of $\boldsymbol{p}_1$, $\boldsymbol{p}_2$ to obtain $\boldsymbol{p}_1',\boldsymbol{p}_2'$;\\
        $\mathcal{P}^{\mathrm{cross}} \gets \mathcal{P}^{\mathrm{cross}} \bigcup \{ \boldsymbol{p}_1', \boldsymbol{p}_2' \}$; \\  
    }

    \tcc{Mutation Operation}
    Set $n_{\mathrm{mut}} = p_{\mathrm{m}}M$; \\
    \For{$k = 1$ \KwTo $n_{\mathrm{mut}}$}{
        Select $\boldsymbol{p}$ from $\mathcal{P}^{\mathrm{temp}}$; \\
        Apply mutation on $(\mathbf{Q}^{(t-1)}, \mathbf{w}^{(t-1)})$ of $\boldsymbol{p}$ to obtain $\boldsymbol{p}'$; \\
        $\mathcal{P}^{\mathrm{mut}} \gets \mathcal{P}^{\mathrm{mut}} \bigcup \boldsymbol{p}'$ ;

    }

    \tcc{Selection}
    $\mathcal{P}^{\mathrm{temp}} \gets \mathcal{P}^{\mathrm{temp}} \cup \mathcal{P}^{\mathrm{cross}} \cup \mathcal{P}^{\mathrm{mut}}$;\\
    Evaluate objectives of all individuals in $\mathcal{P}^{\mathrm{temp}}$;\\
    Perform non-dominated sorting and compute crowding distance;\\
    Refine the best $M$ individuals in $\mathcal{P}^{\mathrm{temp}}$;
    
    \tcc{LLM-guided Parameter Adjustment}
    Calculate SP and $M_3^*$ metrics for the Pareto solutions in $\mathcal{P}^{\mathrm{temp}}$;\\
    Send population data along with SP and M3* metrics to an LLM;\\
    Obtain new $p_{\mathrm{c}}$ and $p_{\mathrm{m}}$ from the LLM;\\
}
Return $\mathbf{Q}^{\textrm{temp}}$ and $\mathbf{w}^{\textrm{temp}}$ of $\mathcal{P}^{\mathrm{temp}}$ as $\mathbf{Q}^{(t)}$ and $\mathbf{w}^{(t)}$.
\end{algorithm}

\par \textbf{2) AO Stage 2: Optimization of $\mathbf{Q}^{(t)}$ and $\mathbf{w}^{(t)}$:} The joint optimization of $\mathbf{Q}^{(t)}$ and $\mathbf{w}^{(t)}$ forms a high-dimensional, continuous, and non-convex problem. Evolutionary computation algorithms are well suited to handle the problem, as their population-based global search enables effective exploration of high-dimensional spaces and facilitates the discovery of optimal or near-optimal solutions. However, traditional evolutionary computation algorithms suffer from two main limitations, which are premature convergence and inefficient static parameter settings. To address these issues, we integrate an LLM with evolutionary computation algorithms to dynamically adjust algorithmic parameters based on population statistics. Specifically, by providing context-aware parameter tuning, the LLM can enhance the performance of evolutionary computation algorithms. Thus, in this stage, we integrate NSGA-II, which is the commonly used evolutionary computation algorithm, with an LLM to achieve the joint optimization of $\mathbf{Q}^{(t)}$ and $\mathbf{w}^{(t)}$ while maintaining $\mathbf{c}^{(t)}$ and $\mathbf{k}^{(t-1)}$ as fixed parameters. The structure of the LLM-guided NSGA-II is presented in Algorithm \ref{algo: NSGA-II} and the corresponding initialization and evolutionary optimization processes are elaborated as follows.

\par \emph{\textbf{During the initialization process}}, the algorithm first sets its operational parameters, including the maximum number of local iterations $T_{\mathrm{local}}$, the crossover probability $p_{\mathrm{c}}$, and the mutation probability $p_{\mathrm{m}}$. Moreover, a temporary population $\mathcal{P}^{\mathrm{temp}}$ is initialized.

\par \emph{\textbf{During the evolutionary process}}, NSGA-II evolves the population through $T_{\mathrm{local}}$ generations, sequentially applying crossover and mutation operations in each generation to produce offspring individuals.

\par In each generation, NSGA-II first performs a crossover operation to generate $n_{\mathrm{cross}} = p_{\mathrm{c}} M$ offspring individuals. Specifically, an offspring population $\mathcal{P}^{\mathrm{cross}}$ is initialized as an empty set. The algorithm then performs $n_{\mathrm{cross}}/2$ crossover operations, where each operation selects two parent solutions $\boldsymbol{p}_1$ and $\boldsymbol{p}_2$ from $\mathcal{P}^{\mathrm{temp}}$. The crossover operations are applied specifically to $\mathbf{Q}^{(t-1)}$ and $\mathbf{w}^{(t-1)}$ of both parents, so that generating two offspring solutions $\boldsymbol{p}_1'$ and $\boldsymbol{p}_2'$. These offspring solutions are then added to the crossover population $\mathcal{P}^{\mathrm{cross}}$.

\par Following crossover, mutation operations are applied to introduce additional diversity into the population. Specifically, the following mutation operation begins by calculating the number of mutation operations as $n_{\mathrm{mut}} = p_{\mathrm{m}}M$. An empty mutation population $\mathcal{P}^{\mathrm{mut}}$ is initialized to store the mutated solutions. For each of the $n_{\mathrm{mut}}$ operations, a parent $\boldsymbol{p}$ is randomly selected from the $\mathcal{P}^{\mathrm{temp}}$. Mutation operations are applied to $\mathbf{Q}^{(t-1)}$ and $\mathbf{w}^{(t-1)}$, so that generating a mutated solution $\boldsymbol{p}'$. The mutated solution is then added to the mutation population $\mathcal{P}^{\mathrm{mut}}$.

\par After generating crossover and mutation offspring, the algorithm performs a selection step to retain high-quality and diverse individuals. The process begins by merging the temporary population with both the crossover and mutation populations, expressed as $\mathcal{P}^{\mathrm{temp}} \gets \mathcal{P}^{\mathrm{temp}} \cup \mathcal{P}^{\mathrm{cross}} \cup \mathcal{P}^{\mathrm{mut}}$. The algorithm then evaluates the objective functions for all solutions in $\mathcal{P}^{\mathrm{temp}}$. Next, all individuals in $\mathcal{P}^{\mathrm{temp}}$ are ranked by non-dominated sorting to identify Pareto fronts. For each front, the crowding distance is computed to preserve population diversity, and dominated individuals are subsequently truncated to maintain a constant population size $M$.

\par After selection, the algorithm introduces LLM to conduct parameter adjustment\footnote{The related code for adjusting algorithm optimization parameters by using LLMs in the MATLAB platform can be found in: \url{https://github.com/xiaoya257248/LLMs_generated-parameters}}. Specifically, this process calculates the SP and $M_3^*$ metrics for the current Pareto front~\cite{cheng2012performance}. These metrics, along with relevant population data, are sent to an LLM. Based on these information, the LLM analyzes the optimization state and generates recommendations to adjust parameter values, i.e., the crossover probability $p_c$ and mutation probability $p_m$. For example, if the SP metric indicates poor distribution of solutions and the $M_3^*$ metric shows limited spread, the LLM might recommend increasing the mutation probability to promote exploration. The response of LLM is parsed to extract the recommended parameter values, which are then used for the next generation of the evolutionary process.

\par Upon completion of all $T_{\mathrm{local}}$ iterations of the evolution process, the algorithm returns $\mathbf{Q}^{(t)}$ and $\mathbf{w}^{(t)}$ of $\mathcal{P}^{\mathrm{temp}}$ as $\mathbf{Q}^{(t)}$ and $\mathbf{w}^{(t)}$. These optimized parameters, along with $\mathbf{c}^{(t)}$ and $\mathbf{k}^{(t-1)}$ are then input to \textbf{AO Stage 3} for the optimization of $\mathbf{k}^{(t)}$ in current iteration.

\begin{algorithm}[htbp]
\caption{GSO}
\label{algo: Greedy strategy for K}

\KwIn{$\mathbf{c}^{(t)}$, $\mathbf{Q}^{(t)}$, $\mathbf{w}^{(t)}$, $\mathbf{k}^{(t-1)}$, $M$, $k_{\mathrm{min}}$, $k_{\mathrm{max}}$.} 

\KwOut{$\mathbf{k}^{(t)}$.}

\For{$m = 1$ \textbf{to} $M$}
{

    $\mathbf{k}^{(t)}_{m} \gets \mathbf{k}^{(t-1)}_{m}$; \\
    
    Calculate $N^{(t)}_{\mathrm{cluster}}$ via Eq.~\eqref{eq. calculation of the number of UAV clusters}; \\
    
    \tcc{Main Optimization}
    \For{$i = 1$ $\textbf{to}$ $N^{(t)}_{\mathrm{cluster}}$}{
    
        $[f_2^{\mathrm{best}}]_{i} = -\infty$;
        $[k^{\mathrm{best}}_m]_{i} = [k^{(t)}_m]_{i}$;
        $\mathbf{k}^{\mathrm{temp}}_{m} \gets \mathbf{k}^{(t)}_{m}$; \\ 
        
        \For{$k = k_{\mathrm{min}}$ $\textbf{to}$ $k_{\mathrm{max}}$}{
            
            Set $[k^{\mathrm{temp}}_m]_{i} = k$; \\
            Calculate $f^{\mathrm{temp}}_2$ based on $\mathbf{c}^{(t)}_{m}$, $\mathbf{Q}^{(t)}_{m}$, $\mathbf{w}^{(t)}_{m}$, and $\mathbf{k}^{\mathrm{temp}}_m$;
                    
            \If{$f^{\mathrm{temp}}_2 > [f_2^{\mathrm{best}}]_{i}$}{
                 $[f_2^{\mathrm{best}}]_{i} = f^{\mathrm{temp}}_2$; \\
                 $[k^{\mathrm{best}}_m]_i = k$; \\ 
            }
        }
        Set $[k^{(t)}_{m}]_i = [k^{\mathrm{best}}_m]_i$;
    }
}
Return $\mathbf{k}^{(t)}$.

\end{algorithm}

\par \textbf{3) AO Stage 3: Optimization of $\mathbf{k}^{(t)}$:} Once $\mathbf{c}^{(t)}$, $\mathbf{Q}^{(t)}$, and $\mathbf{w}^{(t)}$ are fixed, optimizing $\mathbf{k}^{(t)}$ becomes a low-dimensional discrete problem, as clusters can be optimized independently. Compared with dynamic programming, which incurs unnecessary computational overhead, and evolutionary computation, which requires extensive fitness evaluations, a greedy-based strategy can efficiently obtain the optimal solution with lower computational complexity. Thus, we propose GSO to optimize $\mathbf{k}^{(t)}$ in this stage, while keeping $\mathbf{c}^{(t)}$, $\mathbf{Q}^{(t)}$, and $\mathbf{w}^{(t)}$ fixed. \emph{Similarly, the proposed GSO iteratively optimizes the number of semantic symbols per UAV cluster of each individual in the population, i.e., $\mathbf{k}_1^{(t)}, \ldots, \mathbf{k}_M^{(t)}$.} The structure of the proposed GSO is presented in Algorithm \ref{algo: Greedy strategy for K} and the detailed processes are as follows.

\par The GSO initializes the $\mathbf{k}^{(t)}_{m}$ via assigning $\mathbf{k}^{(t-1)}_{m}$ to it and calculates the total number of clusters based on Eq.~\eqref{eq. calculation of the number of UAV clusters}.

\par \emph{Subsequently}, the algorithm optimizes the number of semantic symbols $[k^{(t)}_{m}]_i$ for each cluster $i \in \{1, \ldots, N^{(t)}_{\mathrm{cluster}}\}$, while keeping all other semantic symbols fixed. The process aims to maximize the objective function $f_2$. To this end, several variables are initialized. Specifically, the best benefit of $f_2$, denoted as $[f_2^{\mathrm{best}}]_{i}$, is initialized as $-\infty$ and the corresponding best number of symbols $[k^{\mathrm{best}}_m]_{i}$ is set as $[k^{(t)}_m]_{i}$. Moreover, a temporary vector $\mathbf{k}^{\mathrm{temp}}_{m}$ is defined as a copy of $\mathbf{k}^{\mathrm{(t)}}_{m}$. The GSO then evaluates a candidate integer $k$ for the $i$-th cluster, where $k$ is drawn from the feasible range $[k_{\min}, k_{\max}]$. The $[k^{\mathrm{temp}}_m]_{i}$ is set to $k$, and the resulting second objective function value, $f_2^{\mathrm{temp}}$, is calculated based on the fixed variables ($\mathbf{c}^{(t)}_{m}$, $\mathbf{Q}^{(t)}_{m}$, $\mathbf{w}^{(t)}_{m}$), and $\mathbf{k}^{\mathrm{temp}}_m$. If the calculated $f_2^{\mathrm{temp}}$ exceeds the recorded best benefit $[f_2^{\mathrm{best}}]_{i}$, it can be concluded that $k$ is better than $[k^{\mathrm{best}}_m]_i$. Accordingly, the auxiliary variable $[f_2^{\mathrm{best}}]_{i}$ is defined as $f^{\mathrm{temp}}_2$ and $[k^{\mathrm{best}}_m]_i$ is set as $k$. After evaluating all possible values of $k$, the number of semantic symbols $[k^{(t)}_{m}]_i$ of the $i$-th cluster is updated to its best-performing value $[k^{\mathrm{best}}_m]_i$.

\par \emph{Finally}, GSO returns the optimized $\mathbf{k}^{(t)}$. The updated $\mathbf{k}^{(t)}$ are integrated with the previously optimized $\mathbf{c}^{(t)}$, $\mathbf{Q}^{(t)}$, and $\mathbf{w}^{(t)}$ to form the current population $\mathcal{P}^{(t)}$, which is then input into \textbf{AO Stage 4} for comprehensive assessment.

\par \textbf{4) AO Stage 4: Assessment:} All individuals in $\mathcal{P}^{(t)}$ are first categorized by using non-dominated sorting, followed by the calculation of crowding distance to maintain diversity. \emph{Subsequently}, the population is ranked based on the calculated crowding distance to establish a dominance hierarchy. If the population size $\mathcal{P}^{(t)}$ exceeds the predefined limit $M$, the population is truncated by retaining only the top $M$ individuals according to this dominance hierarchy, which then form the starting population for the subsequent iteration $\mathcal{P}^{(t+1)}$.

\begin{figure*}[htbp]
\centering
\subfigure[2D distribution of UAVs]{
\includegraphics[width=0.45\linewidth]{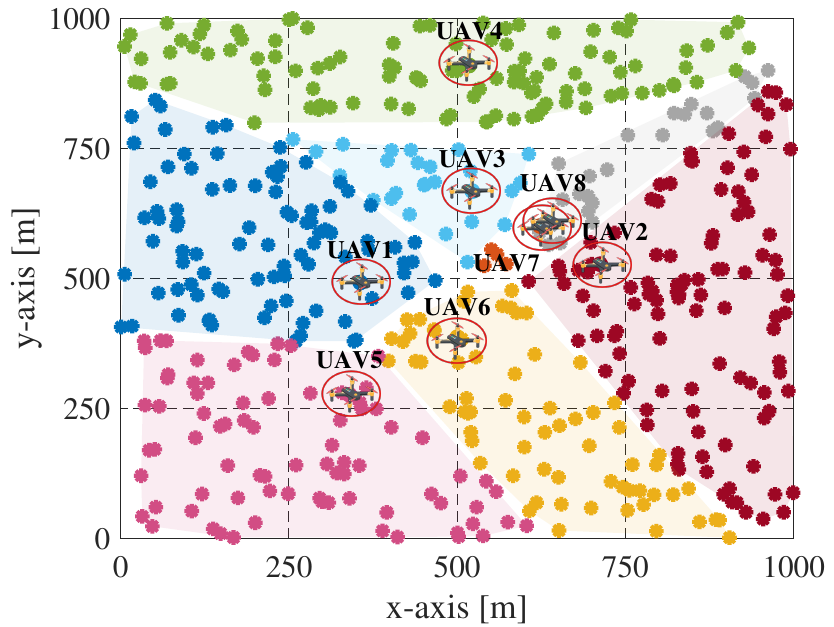}
}
\subfigure[3D distribution of UAVs]{
\includegraphics[width=0.44\linewidth]{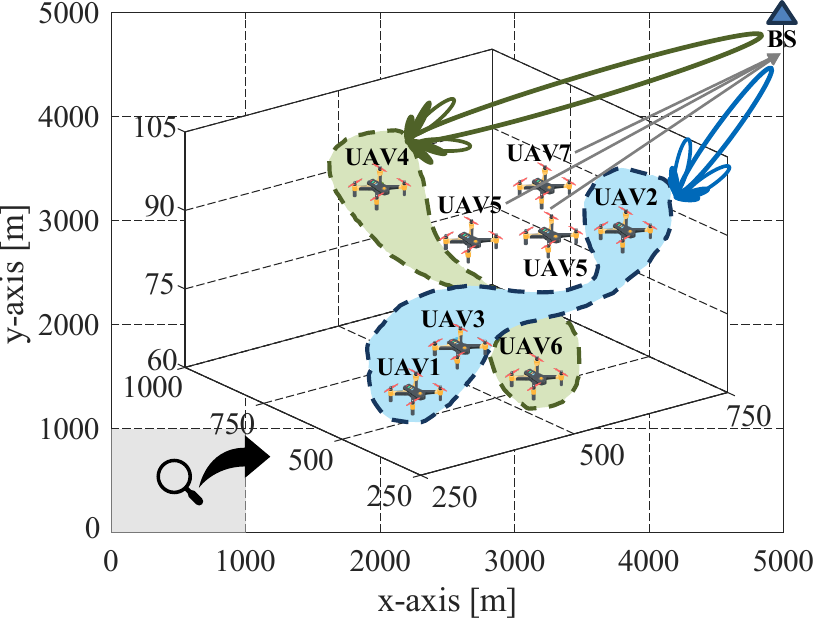}
}
\caption{Spatial distribution of UAVs and the corresponding coverage of ground users within a 1000 m × 1000 m monitor area during conducting data collection.}
\label{fig: Spatial distribution of UAVs and the corresponding coverage of ground users in the 1000 m × 1000 m monitor area when conducting data collection}
\end{figure*}

\subsubsection{Complexity Analysis}
\label{subsec: Complexity Analysis}

\noindent In this part, we analyze the complexity of the proposed LLM-AOA. Specifically, the analysis is conducted from the following three aspects:

\begin{itemize}

\item \emph{Complexity of the GCA:} The while loop can execute a maximum number of $N_{\text{V}} - 1$ iterations before potentially converging to a single cluster. During the $t$-th iteration, let the current number of clusters be $K_t = |\mathscr{C}^{(t)}|$, then the GCA examines $K_t(K_t-1)$ potential cluster pairs for merging. For each merge operation, the GCA performs $N_{\text{V}}$ operations in the worst case for cluster re-indexing. Therefore, the overall computational complexity of the GCA can be expressed as $\mathcal{O}\left((N_{\text{V}}-1) \cdot M \cdot K_t^2 \cdot N_{\text{V}}\right)$. In the worst-case scenario where $K_t$ approaches $N_{\text{V}}$, the complexity of GCA is $\mathcal{O}(M \cdot N_{\text{V}}^4)$.

\item \emph{Complexity of the LLM-guided NSGA-II:} The computational complexity of the LLM-guided NSGA-II is dominated by non-dominated sorting with complexity $\mathcal{O}(M^2)$ and crowding distance calculation with complexity $\mathcal{O}(M\log M)$~\cite{li2023multi}. Since the non-dominated sorting operation has the higher complexity, the overall computational complexity of the LLM-guided NSGA-II is $\mathcal{O}(T_{\mathrm{local}} M^2)$.

\item \emph{Complexity of the GSO:} The complexity of GSO mainly depends on the loop. The intermediate for-loop executes $N^{(t)}_{\mathrm{cluster}}$ iterations and the inner for-loop iterates through all possible values between $k_{\mathrm{min}}$ and $k_{\mathrm{max}}$, resulting in $K_{\text{range}} = k_{\mathrm{max}} - k_{\mathrm{min}} + 1$ iterations. Given that $N^{(t)}_{\mathrm{cluster}}$ upper-bounded by $N_{\mathrm{V}}$ and $K_{\text{range}}$ is a constant, the complexity of the GSO can be written as $\mathcal{O}(M \cdot N_{\mathrm{V}})$.

\end{itemize}

\par In conclusion, the complexity of the proposed LLM-AOA is simplified as $\mathcal{O}\left(T_{\mathrm{max}}^{\mathrm{AO}} \cdot M \cdot (N_{\text{V}}^4 + T_{\mathrm{local}} \cdot M)\right)$. 

%
%
\section{Simulation}
\label{sec: simulation}

\noindent In this section, we present simulation results to illustrate the performance of the proposed LLM-AOA in solving the formulated problem.

\subsection{Simulation Setup}
\label{subsec: Simulation Setup}

\subsubsection{Scenario Setup}
\label{subsubsec: Scenario Setup}

\noindent In this work, the number of ground users $N_{\mathrm{U}}$ and UAVs $N_{\mathrm{V}}$ are set as 500 and 8, respectively. Specifically, the locations of the ground users are defined over a $1000$ m $\times$ 1000 m area and UAVs are constrained to fly at altitudes ranging from 60 m to 120 m above this area. Moreover, the location of the remote BS is set at (5000, 5000, 0) m. 

\subsubsection{Parameter Setup}
\label{subsubsec: Parameter Setup}

\noindent The maximum numbers of AO iterations $T_{\mathrm{max}}^{\mathrm{AO}}$ and LLM-guided NSGA-II $T_{\mathrm{local}}$ are 50 and 10, respectively. The population size $M$ is set as 30 and the initial crossover as well as mutation probabilities are set as 0.8 and 0.4, respectively, which follows common practices in evolutionary computation to balance exploitation and exploration. Moreover, the transmit powers of each ground user $P_{u}$ and each UAV $P_{v}$ are both set as 0.1 w~\cite{zheng2025}. The wavelength $\lambda$, frequency $f$, bandwidth $B$, and noise power density $N_0$ are set to 0.125 m, 2.4 MHz, 2 MHz~\cite{zheng2025}, and -174dBm/Hz~\cite{liu2021uav}, respectively. The lower $k_{\mathrm{min}}$ and upper bound $k_{\mathrm{max}}$ of the average number of semantic symbols are set as 1 and 20~\cite{Hu2025TCOM}, respectively. In addition, the parametersof the channel model, \emph{i.e.}, $\psi$, $\beta$, $\mu_{\mathrm{LoS}}$, and $\mu_{\mathrm{NLoS}}$, are set to 9.61, 0.16, 1.6 dB, and 20 dB, respectively. Finally, the adopted LLM is ChatGPT 5.0.

\subsection{Visualization Results of the Proposed LLM-AOA}
\label{subsec: Visualization Results of the Proposed LLM-AOA}

\noindent Fig. \ref{fig: Spatial distribution of UAVs and the corresponding coverage of ground users in the 1000 m × 1000 m monitor area when conducting data collection} demonstrates the spatial distribution of UAVs and their coverage for ground users in a 1000 m $\times$ 1000 m area. Specifically, different color-coded asterisks represent ground users covered by different UAVs and the labeled UAV icons indicate the placements of all UAVs deployed to optimize coverage. Moreover, UAVs belonging to the same cluster are indicated by circles of the same color, whereas those without circular markers are independent UAVs that do not form clusters with others. 

\par As can be observed, the deployed UAVs exhibit an overall balanced ground user coverage, though UAV 7 is an exception. The pattern arises from the computational capabilities and comparable communication resource limits among the deployed UAVs, under which balanced user coverage ensures uniform resource utilization and mitigates potential bottlenecks due to load imbalance. Moreover, we find that there are two different transmission strategies, which are collaborative and independent transmission, respectively. To be specific, UAVs 1, 2, and 3 form a cluster, while UAVs 4 and 6 constitute another. Within each cluster, data sharing and CB are employed to enhance signal gain and transmission efficiency. In contrast, UAVs 5, 7, and 8 operate separately, which indicates that collaborative transmission does not usually yield superior performance gains for all UAVs. Finally, the spatial distribution of UAVs exhibits two significant characteristics. On the one hand, most UAVs are deployed at or near the geometric centers of their served ground users. Such deployment feature aligns with the min–max distance principle which minimizes the maximum user-to-UAV distance, so that ensuring link quality for edge users and improving the service quality. On the other hand, UAVs engaged in collaborative transmission tend to converge toward a common geometric center, as can be observed from the locations of UAVs 1, 2, and 3. The reason is that relatively compact deployment of collaborative UAVs facilitates higher beamforming gain. The aforementioned spatial distribution of UAVs reflects a trade-off between enhancing the service quality of ground users and improving collaborative transmission efficiency.

\subsection{Comparison of the Proposed LLM-AOA with Other Benchmarks}
\label{subsec: Comparison of the Proposed LLM-AOA with Other Benchmarks}

\noindent We introduce several benchmarks to make comparisons with the proposed LLM-AOA, which are multi-objective differential evolution (MODE)~\cite{babu2003differential}, multi-objective evolutionary algorithm based on decomposition (MOEA/D)~\cite{martinez2014multi}, multi-objective particle swarm optimization (MOPSO)~\cite{MOPSO2008}, NSGA-II, and AOA without the assistance of LLM, respectively.

\par Fig. \ref{fig. pareto solution distribution} illustrates the Pareto solution distributions obtained by the proposed LLM-AOA and other different benchmarks in addressing the formulated DCSFMOP. It is evident that the solutions obtained by the proposed LLM-AOA are closer to the direction of Pareto front (PF), which implies that the performance of LLM-AOA is superior to the other adopted swarm intelligence algorithms. Moreover, the numerical results in Fig. \ref{fig: Numerical Optimization Results Obtained by the proposed LLM-AOA and other benchmarks} further confirm this conclusion since the proposed LLM-AOA achieves the highest transmission rate, \emph{i.e.}, $1.10 \times 10^{8}$ bps, and semantic rate, \emph{i.e.}, $4.12 \times 10^{6}$ suts/s, while maintaining reasonable energy consumption. Under comparable energy consumption levels, LLM-AOA achieves approximately 26.8\% and 22.9\% improvements in the performance of transmission rate and semantic rate, respectively, compared with the conventional AOA. The enhanced performance can be attributed to the fact that the proposed LLM-AOA employs different schemes tailored to the specific characteristics of different components of decision variables, which allows for more effective optimization of each decision variable component and enables LLM-AOA to navigate the complex solution space more efficiently than other algorithms, resulting in better quality solutions.

\begin{figure}
\centerline{\includegraphics[width=3.3in]{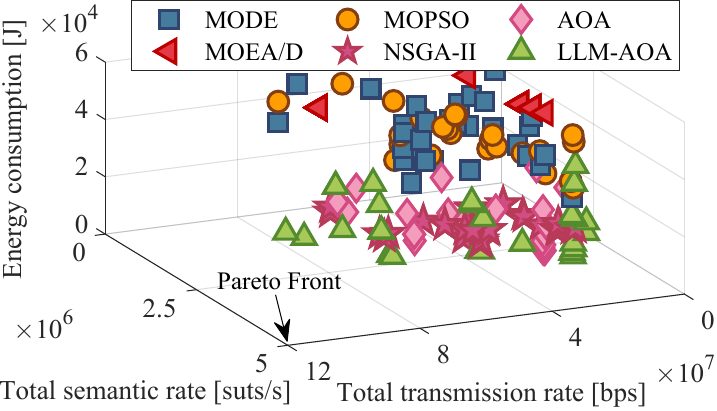}}
\caption{Solution distributions obtained by the proposed LLM-AOA and other benchmarks.}
\label{fig. pareto solution distribution}
\end{figure}

\begin{figure}[htbp]
\centering
\subfigure[f1]{
\includegraphics[width=0.3\linewidth]{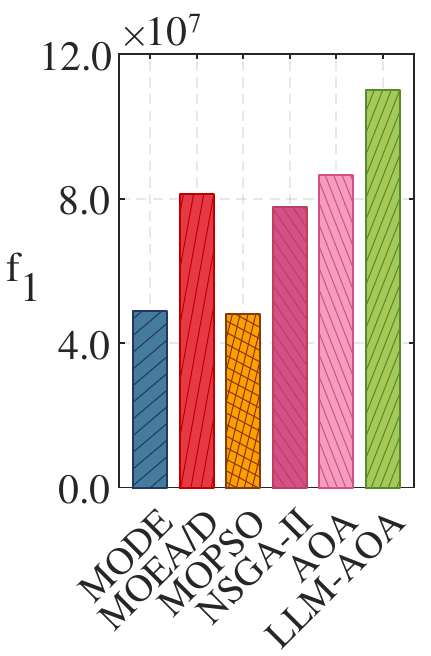}
}
\subfigure[f2]{
\includegraphics[width=0.3\linewidth]{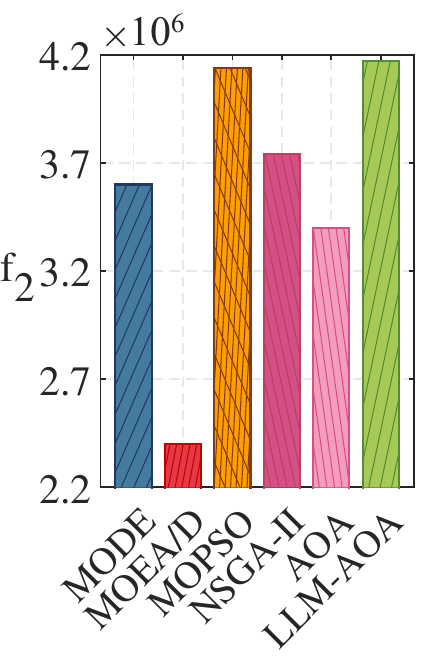}
}
\subfigure[f3]{
\includegraphics[width=0.3\linewidth]{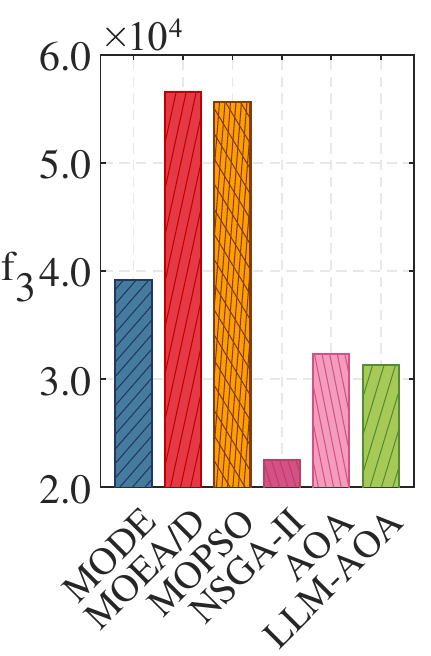}
}
\caption{Numerical Optimization Results Obtained by the proposed LLM-AOA and other benchmarks.}
\label{fig: Numerical Optimization Results Obtained by the proposed LLM-AOA and other benchmarks}
\end{figure}

\subsection{Performance Evaluation of the Proposed LLM-AOA under Different Numbers of UAVs}
\label{subsec: Performance Evaluation under Different Numbers of UAVs}

\noindent To further verify the effectiveness of the proposed LLM-AOA in solving the formulated problem and to elucidate the impact of the number of UAVs on system performance, we extend the simulation by evaluating the performance of the proposed LLM-AOA and other benchmarks under 4, 12, and 16 UAVs, respectively. 

\par Fig. \ref{fig: Optimization results of different objective functions for the proposed LLM-AOA and other benchmarks under different networks sizes} demonstrates the optimization results of different objective functions for the proposed LLM-AOA and other benchmarks under the aforementioned three different networks sizes. As can be observed, although the proposed LLM-AOA yields marginally higher values for $f_{2}$ and $f_{3}$ in certain scenarios, it achieves a more favorable balance among all optimization objectives, which demonstrates its superior optimization capabilities. Moreover, as the number of UAVs increases, the values of $f_{1}$ and $f_{2}$ exhibit an approximately linear growth tendency, which indicates that enlarging the scale of UAV network can substantially enhance the data collection and forwarding capabilities. Conversely, $f_{3}$ exhibits a distinctly different pattern. Specifically, the majority of benchmarks present a substantial increase in energy consumption when the number of UAVs is expanded from 12 to 16 elements, with energy consumption surpassing $2 \times 10^{5}$. Such phenomenon indicates a considerable deterioration in energy efficiency as the network scales beyond certain thresholds. In conclusion, in the considered system, deploying 8 or 12 UAVs offers a more favorable balance among data collection, forwarding performance, and energy consumption, thereby achieving a more desirable trade-off between system performance and energy efficiency.

\begin{figure*}[htbp]
\centering
\subfigure[$f_1$]{
\includegraphics[width=0.3\linewidth]{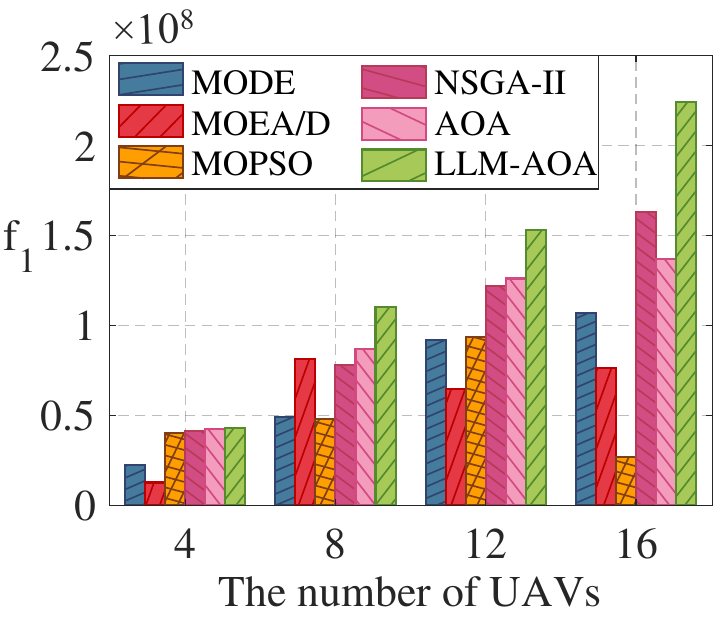}
}
\subfigure[$f_2$]{
\includegraphics[width=0.3\linewidth]{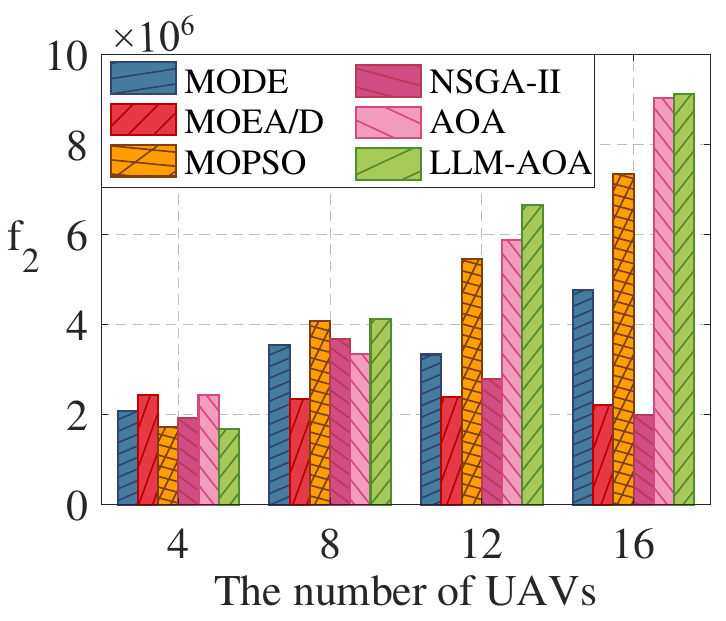}
}
\subfigure[$f_3$]{
\includegraphics[width=0.3\linewidth]{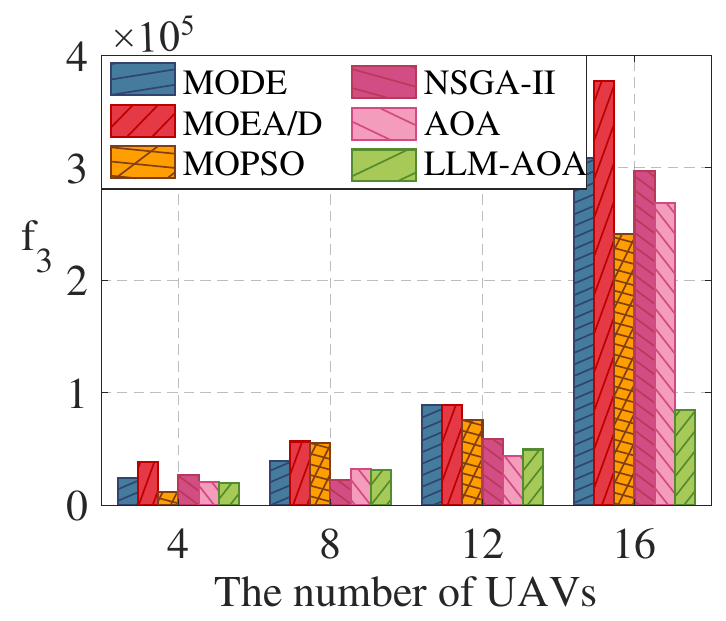}
}
\caption{Optimization results of different objective functions for the proposed LLM-AOA and other benchmarks under different networks sizes.}
\label{fig: Optimization results of different objective functions for the proposed LLM-AOA and other benchmarks under different networks sizes}
\end{figure*}

\subsection{Performance Evaluation of the Proposed LLM-AOA under Different LLMs}
\label{subsec: Performance Evaluation under Different LLMs}

\noindent To evaluate the impact of different LLMs on optimization performance, we conduct extensive simulations by integrating five LLMs into the proposed LLM-AOA, which are ChatGPT, Claude, DeepSeek, Grok, and Qwen, respectively.

\begin{figure}
\centerline{\includegraphics[width=3.3in]{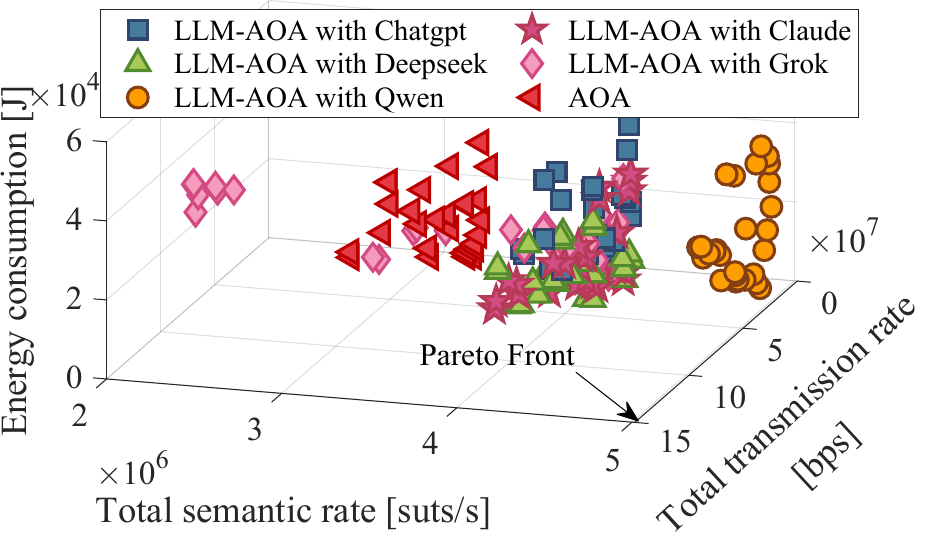}}
\caption{Solution distributions obtained by the proposed LLM-AOA under several different LLMs and AOA.}
\label{fig. pareto solution distribution under different LLMs}
\end{figure}

\par Fig. \ref{fig. pareto solution distribution under different LLMs} illustrates the solution distributions obtained by the proposed LLM-AOA with the aforementioned LLMs and AOA. It can be observed that most LLMs which are integrated into LLM-AOA can deliver favorable performance. The only exception is Grok, for which a small fraction of solutions falls below the performance of the AOA. Moreover, different LLMs exhibit distinct tendencies in balancing the considered multiple optimization objectives. Specifically, LLM-AOA with ChatGPT, Claude, and DeepSeek prefers more balanced parameter adjustments, achieving favorable trade-offs among semantic rate, transmission rate, and energy consumption. In contrast, LLM-AOA with Qwen prioritizes semantic rate, which yields higher semantic rate at the expense of relatively lower ground user–to–UAV transmission rates, which reflects a more aggressive single-objective preference. Finally, the solution set produced by Grok is relatively scattered, with some solutions are superior to the AOA, which suggests its limited effectiveness in handling complex multi-objective trade-offs in the considered scenario.

\section{Conclusion}
\label{sec: Conclusion}

\par In this work, we have designed a low-altitude multi-UAV-assisted data collection and semantic forwarding communication network architecture for post-disaster relief. In the proposed architecture, UAVs first collect data from ground users, adaptively form different clusters, perform intra-cluster data aggregation with semantic extraction, and then cooperate as VAAs to transmit the key data to a remote BS. We have formulated a DCSFMOP, which is a MINLP problem that is inherently NP-hard and characterized by dynamically varying decision variable dimensionality. Considering the inherent characteristics of the formulated problem, we have proposed an LLM-AOA which optimizes different subsets of decision variables through tailored optimization strategies. Simulation results have demonstrated the effectiveness and superiority of the proposed LLM-AOA. Moreover, the solutions obtained by most LLMs achieve favorable performance, except for a subset of solutions obtained by Grok. In addition, LLM-AOA with ChatGPT, Claude, and DeepSeek tends to promote balanced parameter adjustments, thereby yielding well-distributed trade-offs among the three objectives. In contrast, the LLM-AOA with Qwen emphasizes the optimization of a single objective. The aforementioned observations indicate that different LLMs exhibit distinct preferences in balancing multiple optimization objectives.

\par In the future, we aim to extend this research by investigating solar-powered UAV clustering in dynamic environments, exploring multi-modal semantic-driven self-organizing cluster formation, and developing distributed collaboration frameworks for heterogeneous UAV semantic communication networks to further enhance system robustness and efficiency in complex post-disaster scenarios.

\bibliographystyle{IEEEtran}
\bibliography{myref}

\begin{IEEEbiography}[{\includegraphics[width=1in,height=1.25in,clip,keepaspectratio]{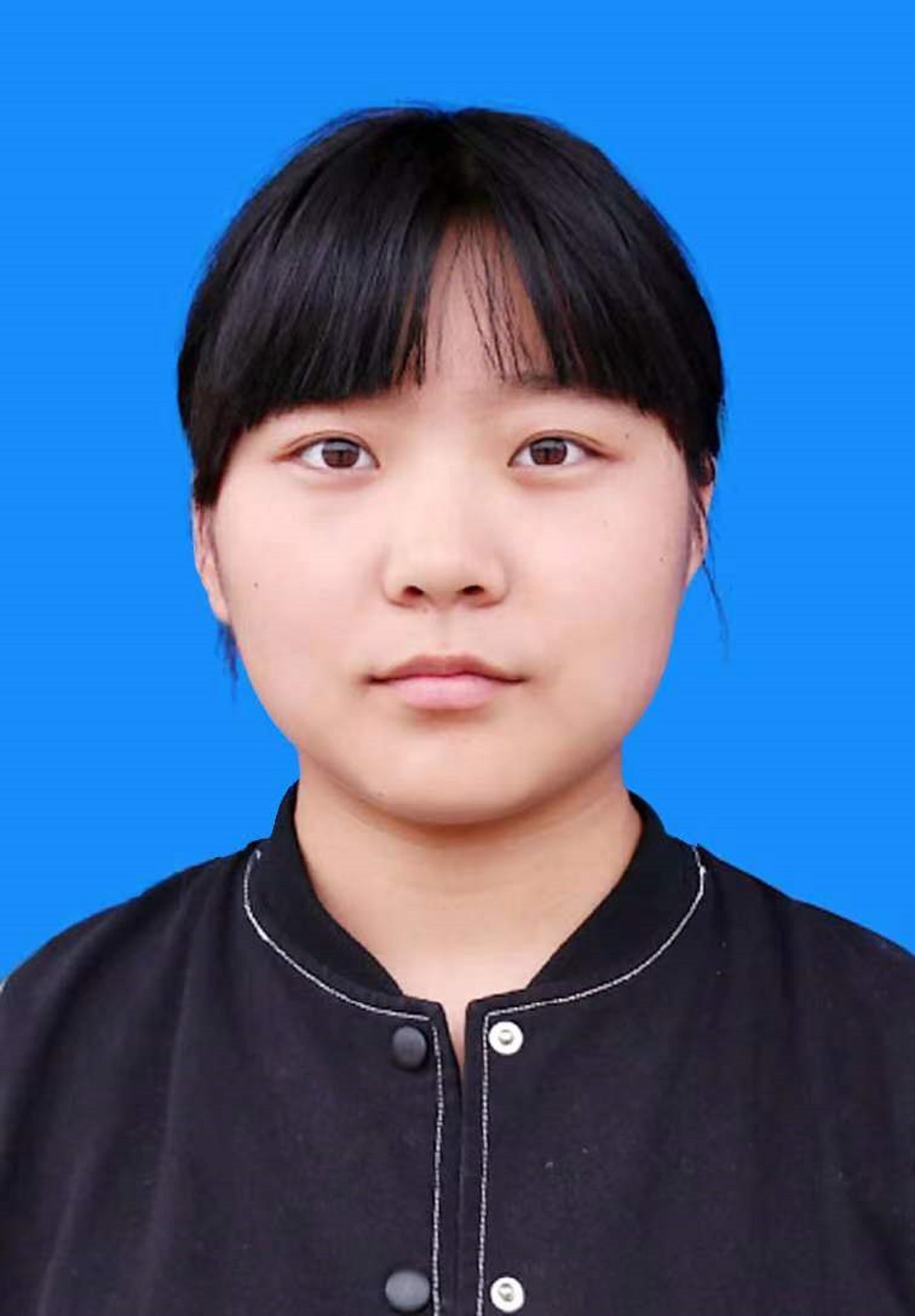}}]{Xiaoya Zheng} received the B.S. degree in software engineering from Hebei Geology University in 2021, and the M.S. degree in College of Computer Science and Technology from Jilin University in 2024. She is currently pursuing a Ph.D. degree at the College of Computer Science and Technology, Jilin University. Her research interests focus on UAV networks and optimization techniques.
\end{IEEEbiography}

\vspace{-10mm}
\begin{IEEEbiography}[{\includegraphics[width=1in,height=1.25in,clip,keepaspectratio]{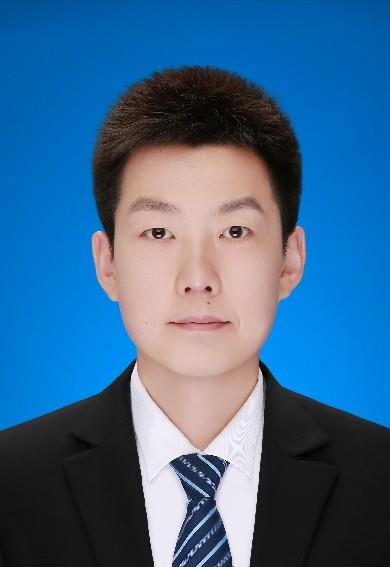}}]{Geng Sun} (Senior Member, IEEE) received the B.S. degree in communication engineering from Dalian Polytechnic University, and the Ph.D. degree in computer science and technology from Jilin University, in 2011 and 2018, respectively. He was a Visiting Researcher with the School of Electrical and Computer Engineering, Georgia Institute of Technology, USA. He is a Professor in the College of Computer Science and Technology at Jilin University. Currently, he is working as a visiting scholar at the College of Computing and Data Science, Nanyang Technological University, Singapore. He has published over 100 high-quality papers, including IEEE TMC, IEEE JSAC, IEEE/ACM ToN, IEEE TWC, IEEE TCOM, IEEE TAP, IEEE IoT-J, IEEE TIM, IEEE INFOCOM, IEEE GLOBECOM, and IEEE ICC. He serves as the Associate Editors of IEEE Communications Surveys \& Tutorials, IEEE Transactions on Communications, IEEE Transactions on Vehicular Technology, IEEE Transactions on Network Science and Engineering, IEEE Transactions on Network and Service Management and IEEE Networking Letters. He serves as the Lead Guest Editor of Special Issues for IEEE Transactions on Network Science and Engineering, IEEE Internet of Things Journal, IEEE Networking Letters. He also serves as the Guest Editor of Special Issues for IEEE Transactions on Services Computing, IEEE Communications Magazine, and IEEE Open Journal of the Communications Society. His research interests include Low-altitude Wireless Networks, UAV communications and Networking, Mobile Edge Computing (MEC), Intelligent Reflecting Surface (IRS), Generative AI and Agentic AI, and deep reinforcement learning.
\end{IEEEbiography}

\vspace{-10mm}
\begin{IEEEbiography}[{\includegraphics[width=1in,height=1.25in,clip,keepaspectratio]{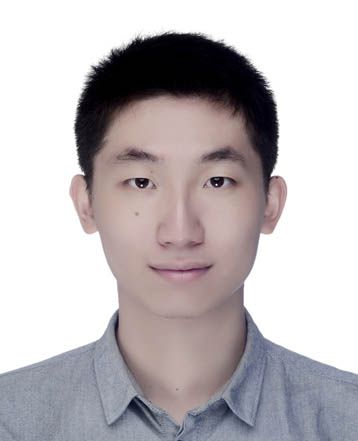}}]{Jiahui Li} (Member, IEEE) received his B.S. in Software Engineering, and M.S. and Ph.D. in Computer Science and Technology from Jilin University, Changchun, China, in 2018, 2021, and 2024, respectively. He was a visiting Ph.D. student at the Singapore University of Technology and Design (SUTD). He currently serves as an assistant researcher in the College of Computer Science and Technology at Jilin University. His current research focuses on integrated air-ground networks, UAV networks, wireless energy transfer, and optimization.
\end{IEEEbiography}

\vspace{-10mm}
\begin{IEEEbiography}[{\includegraphics[width=1in,height=1.25in,clip,keepaspectratio]{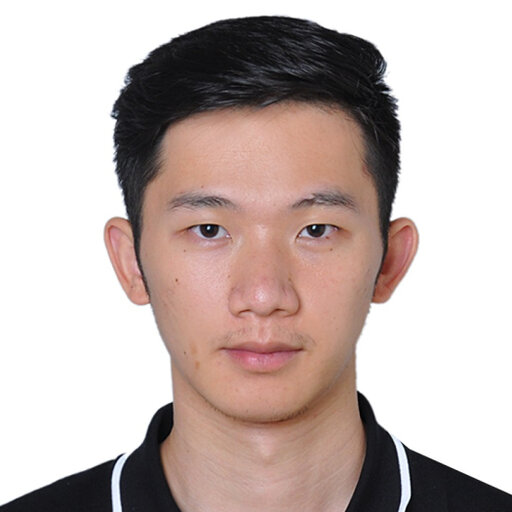}}]{Jiacheng Wang} is the research fellow in the College of Computing and Data Science at Nanyang Technological University, Singapore. Prior to that, he received the Ph.D. degree in School of Communications and Information Engineering, Chongqing University of Posts and Telecommunications, Chongqing, China. His research interests include wireless sensing, semantic communications, and generative AI, Metaverse.
\end{IEEEbiography}

\vspace{-10mm}
\begin{IEEEbiography}[{\includegraphics[width=1in,height=1.25in,clip,keepaspectratio]{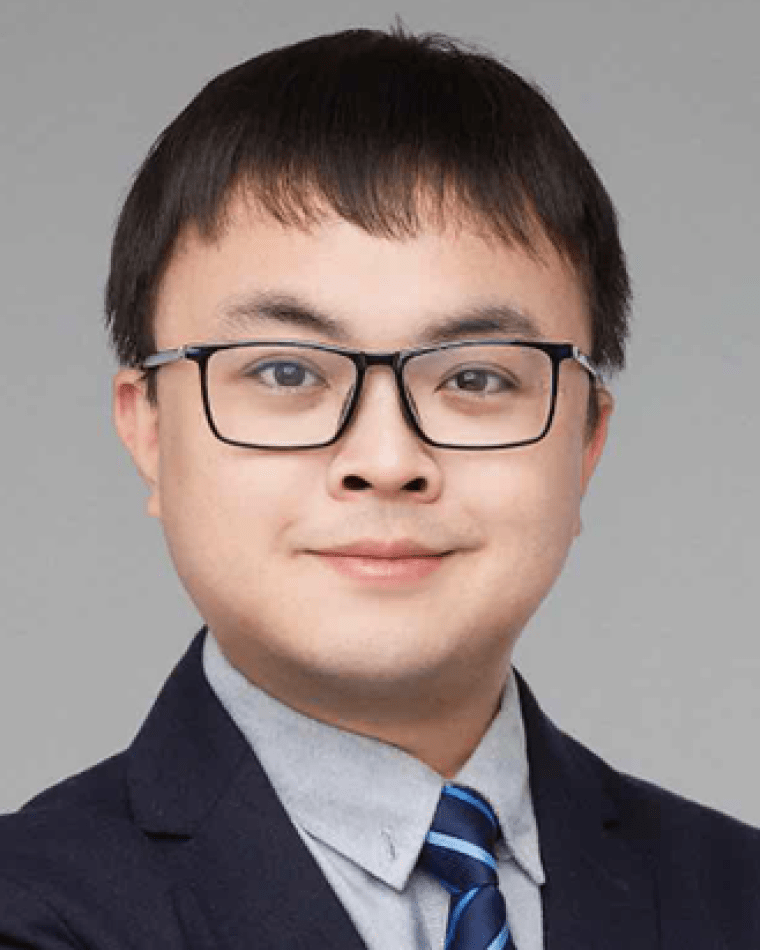}}]{Weijie Yuan} (Senior Member, IEEE) is currently an Assistant Professor with the Southern University of Science and Technology. His research interests include integrated sensing and communications (ISAC), orthogonal time frequency space (OTFS), and low-altitude wireless networks (LAWN). He was a recipient of the Best Editor from IEEE CommL; the Best Paper Award from IEEE ICC 2023, IEEE/CIC ICCC 2023, and IEEE GlobeCom 2024; and the 2025 IEEE Communications Society and Information Theory Society Joint Paper Award. He was the Track-Chair of IEEE ICC 2025 and IEEE VTC 2025-Spring. He served as an Organizer/the Chair for several workshops and special sessions in flagship IEEE and ACM conferences, including IEEE ICC, IEEE VTC, IEEE GlobeCom, IEEE/CIC ICCC, IEEE SPAWC, IEEE WCNC, IEEE ICASSP, and ACM MobiCom. He is the Founding Chair of the IEEE ComSoc Special Interest Group (SIG) on LAWN and the SIG on OTFS. He serves as an Editor for IEEE Transactions on Communications, IEEE Transactions on Wireless Communications, IEEE Transactions on Mobile Computing, IEEE Communications Magazine.
\end{IEEEbiography}

\vspace{-10mm}
\begin{IEEEbiography}[{\includegraphics[width=1in,height=1.25in,clip,keepaspectratio]{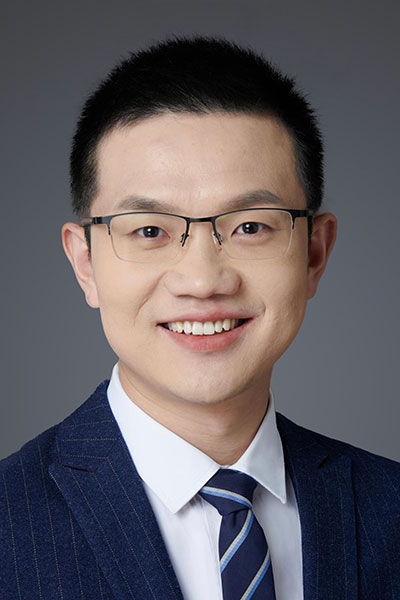}}]{Qingqing Wu} (Senior Member, IEEE) is currently an Associate Professor with Shanghai Jiao Tong University. He has co-authored more than 100 IEEE journal articles with 29 ESI highly cited articles and nine ESI hot articles, which have received more than 20000 Google citations. His current research interest includes intelligent reflecting surface (IRS), uncrewed aerial vehicle (UAV) communications, and MIMO transceiver design.,Dr. Wu was a recipient of the IEEE Communications Society Fred Ellersick Prize, the IEEE Best Tutorial Paper Award in 2023, Asia–Pacific Best Young Researcher Award and Outstanding Paper Award in 2022, the Young Author Best Paper Award in 2021, the Outstanding Ph.D. Thesis Award of China Institute of Communications in 2017, the IEEE ICCC Best Paper Award in 2021, and the IEEE WCSP Best Paper Award in 2015. He is the Workshop Co-Chair of IEEE ICC 2019-2023 and IEEE GLOBECOM 2020. He was the Workshops and Symposia Officer of the Reconfigurable Intelligent Surfaces Emerging Technology Initiative and a Research Blog Officer of the Aerial Communications Emerging Technology Initiative. He is the IEEE Communications Society Young Professional Chair in Asia Pacific Region. He was the Exemplary Editor of IEEE Communications Letters in 2019 and the exemplary reviewer of several IEEE journals. He serves as an Associate Editor for IEEE Transactions on Communications, IEEE Communications Letters, and IEEE Wireless Communications Letters. He is the Lead Guest Editor of IEEE Journal on Selected Areas in Communications. He was listed as the Clarivate ESI Highly Cited Researcher in 2022 and 2021, the Most Influential Scholar Award in AI-2000 by Aminer in 2021, and the World’s Top 2.
\end{IEEEbiography}

\vspace{-10mm}
\begin{IEEEbiography}[{\includegraphics[width=1in,height=1.25in,clip,keepaspectratio]{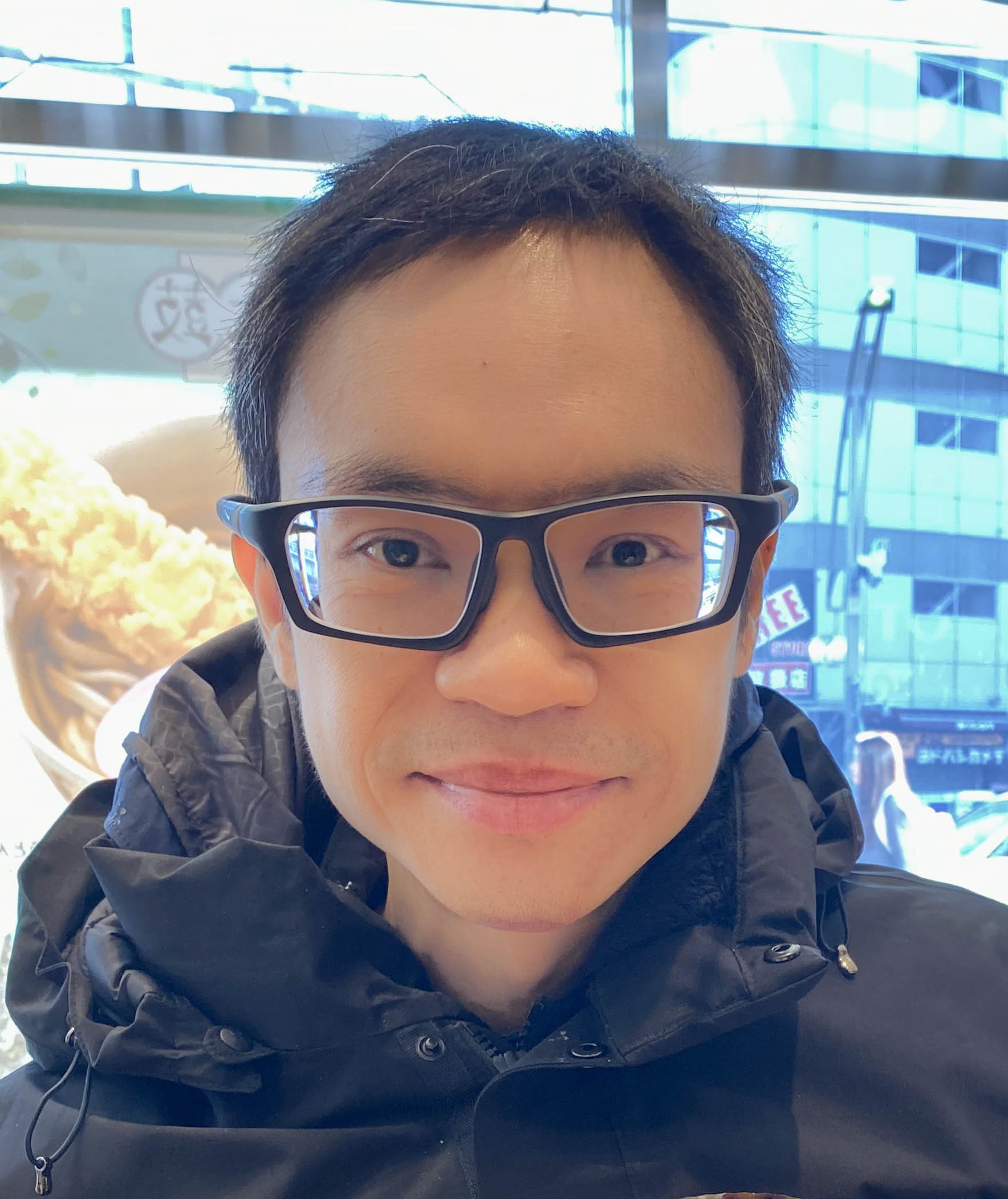}}]{Dusit Niyato} (Fellow, IEEE) is a professor in the College of Computing and Data Science, at Nanyang Technological University, Singapore. He received B.Eng. from King Mongkuts Institute of Technology Ladkrabang (KMITL), Thailand and Ph.D. in Electrical and Computer Engineering from the University of Manitoba, Canada. His research interests are in the areas of mobile generative AI, edge general intelligence, quantum computing and networking, and incentive mechanism design.
\end{IEEEbiography}

\vspace{-10mm}
\begin{IEEEbiography}[{\includegraphics[width=1in,height=1.25in,clip,keepaspectratio]{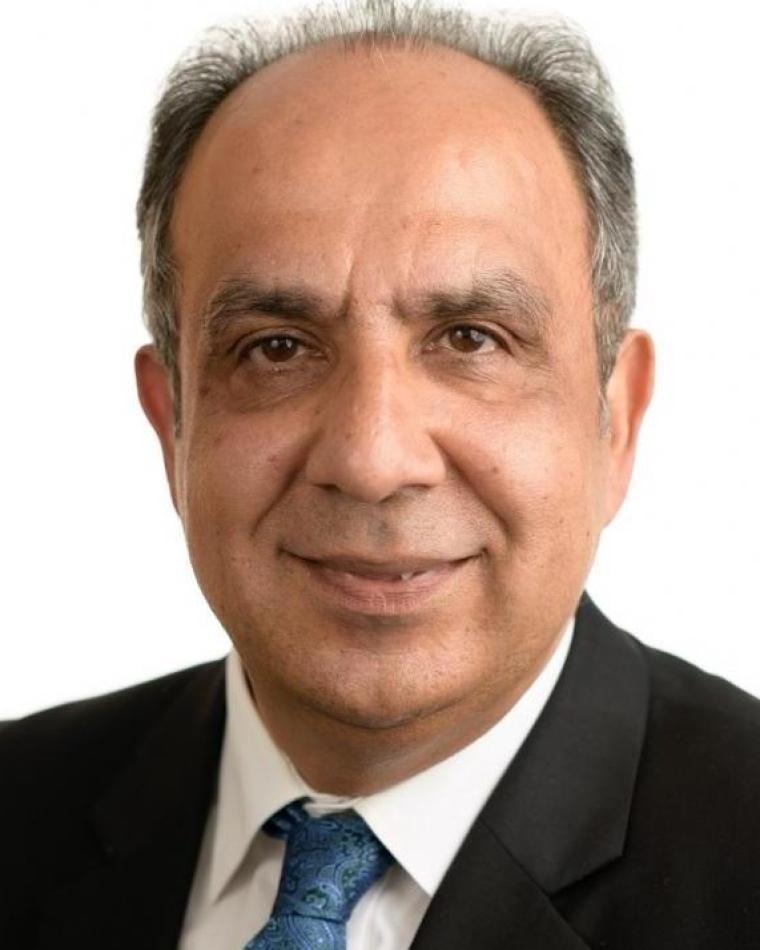}}]{Abbas Jamalipour}
(Fellow, IEEE) received the Ph.D. degree in Electrical Engineering from Nagoya University, Nagoya, Japan in 1996. He is the Professor of Ubiquitous Mobile Networking at the University of Sydney. He has authored nine technical books, eleven book chapters, over 650 technical papers, and five patents, all in wireless communications and networking. Prof. Jamalipour is the recipient of several prestigious awards such as the 2025 Neville Thiele Eminence Award from Engineers Australia, IEEE ComSoc Harold Sobol Award, the IEEE ComSoc Best Tutorial Paper Award, as well as over fifteen Best Paper Awards. He has held positions of President, Executive Vice-President, elected member of the Board of Governors of IEEE Vehicular Technology Society, and the Editor-in-Chief of IEEE TRANSACTIONS ON VEHICULAR TECHNOLOGY and VTS Mobile World. He was the Editor-in-Chief IEEE WIRELESS COMMUNICATIONS, the Vice President-Conferences, and a member of Board of Governors of the IEEE Communications Society, currently he is the Vice President-MGA. He is a Fellow of the Institute of Electrical, Information, and Communication Engineers (IEICE), the Institution of Engineers Australia (IEAust), and the International Artificial Intelligence Industry Alliance (AIIA), and a Visiting Fellow of the UK Royal Academy of Engineering.
\end{IEEEbiography}

\end{document}